\documentclass[a4paper,5pt]{article}
\usepackage[maxbibnames=99]{biblatex}
\usepackage[text={180mm,240mm},centering]{geometry}
 
\usepackage{graphics,graphicx}
\usepackage{subfig}
\usepackage{float}
\usepackage{booktabs}
\usepackage{longtable}
\usepackage{supertabular}
\usepackage{tabularx}
\usepackage{multirow}
\usepackage{array}
\usepackage{tabu}
\usepackage{mathtools}
\usepackage{hyperref}
\usepackage{accents} 

\usepackage[ruled,vlined]{algorithm2e}

\usepackage{amsmath}
\usepackage{amsfonts}
\usepackage{amssymb}
\usepackage{siunitx} 

\usepackage{enumerate}
\usepackage{sectsty}
\usepackage[affil-it,auth-sc]{authblk}
\usepackage{color}
\usepackage[dvipsnames]{xcolor}

\sectionfont{\normalsize}
\subsectionfont{\normalsize}

\usepackage{tikz}
\usetikzlibrary{arrows,decorations,backgrounds,shapes, snakes}
\usetikzlibrary{positioning}
\usetikzlibrary{matrix} 
\usepackage{varwidth}
\usepackage[most]{tcolorbox}
\usetikzlibrary{shapes.geometric,arrows.meta,decorations.markings}

\usepackage{fancyhdr}            
\fancypagestyle{plain}{%
  \fancyhead{}                                     
  \fancyfoot[CE,CO]{\thepage}       
  \fancyhead[CE,CO]{\textcolor[rgb]{0.64,0.15,0.15}{Published in  \emph{Journal of the Mechanics and Physics of Solids} (2023), in press\\
  doi: https://doi.org/10.1016/j.jmps.2023.105452}}
\setlength{\headheight}{14.5pt}}

\begin{document}

\newcommand{\singlespace}{\baselineskip=12pt\lineskiplimit=0pt\lineskip=0pt}
\def\ds{\displaystyle}

\tikzstyle{every picture}+=[remember picture]

\newcommand{\beq}{\begin{equation}}
\newcommand{\eeq}{\end{equation}}
\newcommand{\lb}{\label}
\newcommand{\ph}{\phantom}
\newcommand{\beqar}{\begin{eqnarray}}
\newcommand{\eeqar}{\end{eqnarray}}
\newcommand{\barr}{\begin{array}}
\newcommand{\earr}{\end{array}}
\newcommand{\jump}{\parallel}
\newcommand{\Ehat}{\hat{E}}
\newcommand{\That}{\hat{\bf T}}
\newcommand{\Ahat}{\hat{A}}
\newcommand{\chat}{\hat{c}}
\newcommand{\shat}{\hat{s}}
\newcommand{\khat}{\hat{k}}
\newcommand{\muhat}{\hat{\mu}}
\newcommand{\mc}{M^{\scriptscriptstyle C}}
\newcommand{\mei}{M^{\scriptscriptstyle M,EI}}
\newcommand{\mec}{M^{\scriptscriptstyle M,EC}}
\newcommand{\hbeta}{{\hat{\beta}}}
\newcommand{\rec}[2]{\left( #1 #2 \ds{\frac{1}{#1}}\right)}
\newcommand{\rep}[2]{\left( {#1}^2 #2 \ds{\frac{1}{{#1}^2}}\right)}
\newcommand{\derp}[2]{\ds{\frac {\partial #1}{\partial #2}}}
\newcommand{\derpn}[3]{\ds{\frac {\partial^{#3}#1}{\partial #2^{#3}}}}
\newcommand{\dert}[2]{\ds{\frac {d #1}{d #2}}}
\newcommand{\dertn}[3]{\ds{\frac {d^{#3} #1}{d #2^{#3}}}}
\newcommand{\ct}{\captionof{table}}
\newcommand{\cf}{\captionof{figure}}

\def\c{{\circ}}
\def\bob{{\, \underline{\overline{\otimes}} \,}}
\def\ob{{\, \underline{\otimes} \,}}
\def\scalp{\mbox{\boldmath$\, \cdot \, $}}
\def\gdp{\makebox{\raisebox{-.215ex}{$\Box$}\hspace{-.778em}$\times$}}
\def\daa{\makebox{\raisebox{-.050ex}{$-$}\hspace{-.550em}$: ~$}}
\def\mK{\mbox{${\mathcal{K}}$}}
\def\cK{\mbox{${\mathbb {K}}$}}

\def\Xint#1{\mathchoice
   {\XXint\displaystyle\textstyle{#1}}%
   {\XXint\textstyle\scriptstyle{#1}}%
   {\XXint\scriptstyle\scriptscriptstyle{#1}}%
   {\XXint\scriptscriptstyle\scriptscriptstyle{#1}}%
   \!\int}
\def\XXint#1#2#3{{\setbox0=\hbox{$#1{#2#3}{\int}$}
     \vcenter{\hbox{$#2#3$}}\kern-.5\wd0}}
\def\ddashint{\Xint=}
\def\fpint{\Xint=}
\def\dashint{\Xint-}
\def\cpvint{\Xint-}
\def\intl{\int\limits}
\def\cpvintl{\cpvint\limits}
\def\fpintl{\fpint\limits}
\def\ointl{\oint\limits}
\def\bA{{\bf A}}
\def\ba{{\bf a}}
\def\bB{{\bf B}}
\def\bb{{\bf b}}
\def\bc{{\bf c}}
\def\bC{{\bf C}}
\def\bD{{\bf D}}
\def\bE{{\bf E}}
\def\be{{\bf e}}
\def\bbf{{\bf f}}
\def\bF{{\bf F}}
\def\bG{{\bf G}}
\def\bg{{\bf g}}
\def\bi{{\bf i}}
\def\bI{{\bf I}}
\def\bH{{\bf H}}
\def\bK{{\bf K}}
\def\bL{{\bf L}}
\def\bM{{\bf M}}
\def\bN{{\bf N}}
\def\bn{{\bf n}}
\def\bm{{\bf m}}
\def\b0{{\bf 0}}
\def\bo{{\bf o}}
\def\bX{{\bf X}}
\def\bx{{\bf x}}
\def\bP{{\bf P}}
\def\bp{{\bf p}}
\def\bQ{{\bf Q}}
\def\bq{{\bf q}}
\def\bR{{\bf R}}
\def\bS{{\bf S}}
\def\bs{{\bf s}}
\def\bT{{\bf T}}
\def\bt{{\bf t}}
\def\bU{{\bf U}}
\def\bu{{\bf u}}
\def\bv{{\bf v}}
\def\bV{{\bf V}}
\def\bw{{\bf w}}
\def\bW{{\bf W}}
\def\by{{\bf y}}
\def\bz{{\bf z}}
\def\T{{\bf T}}
\def\Te{\textrm{T}}
\def\Id{{\bf I}}
\def\bxi{\mbox{\boldmath${\xi}$}}
\def\balpha{\mbox{\boldmath${\alpha}$}}
\def\bbeta{\mbox{\boldmath${\beta}$}}
\def\bepsilon{\mbox{\boldmath${\epsilon}$}}
\def\bvarepsilon{\mbox{\boldmath${\varepsilon}$}}
\def\bomega{\mbox{\boldmath${\omega}$}}
\def\bphi{\mbox{\boldmath${\phi}$}}
\def\bsigma{\mbox{\boldmath${\sigma}$}}
\def\bfeta{\mbox{\boldmath${\eta}$}}
\def\bDelta{\mbox{\boldmath${\Delta}$}}
\def\btau{\mbox{\boldmath $\tau$}}
\def\tr{{\rm tr}}
\def\dev{{\rm dev}}
\def\div{{\rm div}}
\def\Div{{\rm Div}}
\def\Grad{{\rm Grad}}
\def\grad{{\rm grad}}
\def\Lin{{\rm Lin}}
\def\Sym{{\rm Sym}}
\def\Skw{{\rm Skew}}
\def\abs{{\rm abs}}
\def\Re{{\rm Re}}
\def\Im{{\rm Im}}
\def\capB{\mbox{\boldmath${\mathsf B}$}}
\def\capC{\mbox{\boldmath${\mathsf C}$}}
\def\capD{\mbox{\boldmath${\mathsf D}$}}
\def\capE{\mbox{\boldmath${\mathsf E}$}}
\def\capG{\mbox{\boldmath${\mathsf G}$}}
\def\tcapG{\tilde{\capG}}
\def\capH{\mbox{\boldmath${\mathsf H}$}}
\def\capK{\mbox{\boldmath${\mathsf K}$}}
\def\capL{\mbox{\boldmath${\mathsf L}$}}
\def\capM{\mbox{\boldmath${\mathsf M}$}}
\def\capR{\mbox{\boldmath${\mathsf R}$}}
\def\capW{\mbox{\boldmath${\mathsf W}$}}

\def\i{\mbox{${\mathrm i}$}}
\def\mC{\mbox{\boldmath${\mathcal C}$}}
\def\mB{\mbox{${\mathcal B}$}}
\def\mE{\mbox{${\mathcal{E}}$}}
\def\mL{\mbox{${\mathcal{L}}$}}
\def\mK{\mbox{${\mathcal{K}}$}}
\def\mV{\mbox{${\mathcal{V}}$}}
\def\C{\mbox{\boldmath${\mathcal C}$}}
\def\E{\mbox{\boldmath${\mathcal E}$}}

\def\AAM{{\it Advances in Applied Mechanics }}
\def\ACME{{\it Arch. Comput. Meth. Engng.}}
\def\ARMA{{\it Arch. Rat. Mech. Analysis}}
\def\AMR{{\it Appl. Mech. Rev.}}
\def\ASCEEM{{\it ASCE J. Eng. Mech.}}
\def\ACTA{{\it Acta Mater.}}
\def\CMAME {{\it Comput. Meth. Appl. Mech. Engrg.}}
\def\CRAS{{\it C. R. Acad. Sci. Paris}}
\def\CRM{{\it Comptes Rendus M\'ecanique}}
\def\EFM{{\it Eng. Fracture Mechanics}}
\def\EJMA{{\it Eur.~J.~Mechanics-A/Solids}}
\def\IJES{{\it Int. J. Eng. Sci.}}
\def\IJF{{\it Int. J. Fracture}}
\def\IJMS{{\it Int. J. Mech. Sci.}}
\def\IJNAMG{{\it Int. J. Numer. Anal. Meth. Geomech.}}
\def\IJP{{\it Int. J. Plasticity}}
\def\IJSS{{\it Int. J. Solids Structures}}
\def\IngA{{\it Ing. Archiv}}
\def\JAM{{\it J. Appl. Mech.}}
\def\JAP{{\it J. Appl. Phys.}}
\def\JAE{{\it J. Aerospace Eng.}}
\def\JE{{\it J. Elasticity}}
\def\JM{{\it J. de M\'ecanique}}
\def\JMPS{{\it J. Mech. Phys. Solids}}
\def\JSV{{\it J. Sound and Vibration}}
\def\MACRO{{\it Macromolecules}}
\def\MMT{{\it Mech. Mach. Th.}}
\def\MOM{{\it Mech. Materials}}
\def\MMS{{\it Math. Mech. Solids}}
\def\MMT{{\it Metall. Mater. Trans. A}}
\def\MPCPS{{\it Math. Proc. Camb. Phil. Soc.}}
\def\MSE{{\it Mater. Sci. Eng.}}
\def\NATURE{{\it Nature}}
\def\NATUREM{{\it Nature Mater.}}
\def\PHIL{{\it Phil. Trans. R. Soc.}}
\def\PMPS{{\it Proc. Math. Phys. Soc.}}
\def\PNAS{{\it Proc. Nat. Acad. Sci.}}
\def\PRE{{\it Phys. Rev. E}}
\def\PRL{{\it Phys. Rev. Letters}}
\def\PRSL{{\it Proc. R. Soc.}}
\def\RIIT{{\it Rozprawy Inzynierskie - Engineering Transactions}}
\def\ROCK{{\it Rock Mech. and Rock Eng.}}
\def\QAM{{\it Quart. Appl. Math.}}
\def\QJMAM{{\it Quart. J. Mech. Appl. Math.}}
\def\SCIENCE{{\it Science}}
\def\SCRMAT{{\it Scripta Mater.}}
\def\SM{{\it Scripta Metall.}}
\def\ZAMM{{\it Z. Angew. Math. Mech.}}
\def\ZAMP{{\it Z. Angew. Math. Phys.}}
\def\ZVDI{{\it Z. Verein. Deut. Ing.}}

\def\salto#1#2{
[\mbox{\hspace{-#1em}}[#2]\mbox{\hspace{-#1em}}]}

\renewcommand\Affilfont{\itshape}
\setlength{\affilsep}{1em}
\renewcommand\Authsep{, }
\renewcommand\Authand{ and }
\renewcommand\Authands{ and }
\setcounter{Maxaffil}{2}

\title{
Stabilization against gravity and self-tuning of 
an elastic variable-length rod\\ through  an oscillating  sliding sleeve
}
\author[1]{P. Koutsogiannakis}
\author[1]{D. Misseroni}
\author[1]{D. Bigoni}
\author[1]{F. Dal Corso\footnote{Corresponding author: francesco.dalcorso@unitn.it}}
\affil[1]{DICAM, University of Trento, via~Mesiano~77, I-38123 Trento, Italy}

\maketitle

\begin{abstract}

An elastic rod, straight in its undeformed state, has a mass attached at one end and a variable length, due to a constraint at the other end by a frictionless sliding sleeve. The constraint is arranged with the sliding direction parallel to a gravity field, in a way that the rod can freely slip inside of the sleeve, when the latter is not moving. In this case, the free fall of the mass continues until the rod is completely injected into the constraint. However, when the sliding sleeve is subject to a harmonic transverse vibration, it is shown that the fall of the mass and the rod injection are  hindered by the presence of a configurational force developing at the sliding sleeve and acting oppositely to gravity. During the dynamic motion, such a configurational force is varying in time because it is  associated with  the variable bending moment at the sleeve entrance. It is  (experimentally, analytically, and numerically) demonstrated that, in addition to the states of
complete injection or ejection of the elastic rod (for which  the mass falls down or is thrown out),  
a stable sustained  
oscillation around a finite height can be realized. This 
\lq suspended motion' is the signature of a new attractor, that arises by the constraint oscillation. 
This behaviour  
shares similarities with  
parametric  oscillators, as for instance 
the Kapitza inverted pendulum. However, differently from the classical parametric oscillators, the \lq suspended' configuration of 
the rod violates  equilibrium and the stabilization occurs through a transverse mechanical input, instead of a longitudinal one. By varying the sliding sleeve oscillation amplitude and frequency within specific sets of values, the system spontaneously adjusts the sustained motion   through a self-tuning of the rod's external length. This self-tuning property    opens the way to  the design of vibration-based devices with extended frequency range. 

\end{abstract}

\vspace{5 mm}

\section{Introduction}

\begin{figure}[!h]
    \renewcommand{\figurename}{\footnotesize{Figure}}
    \begin{center}
    
       \includegraphics[width=140mm]{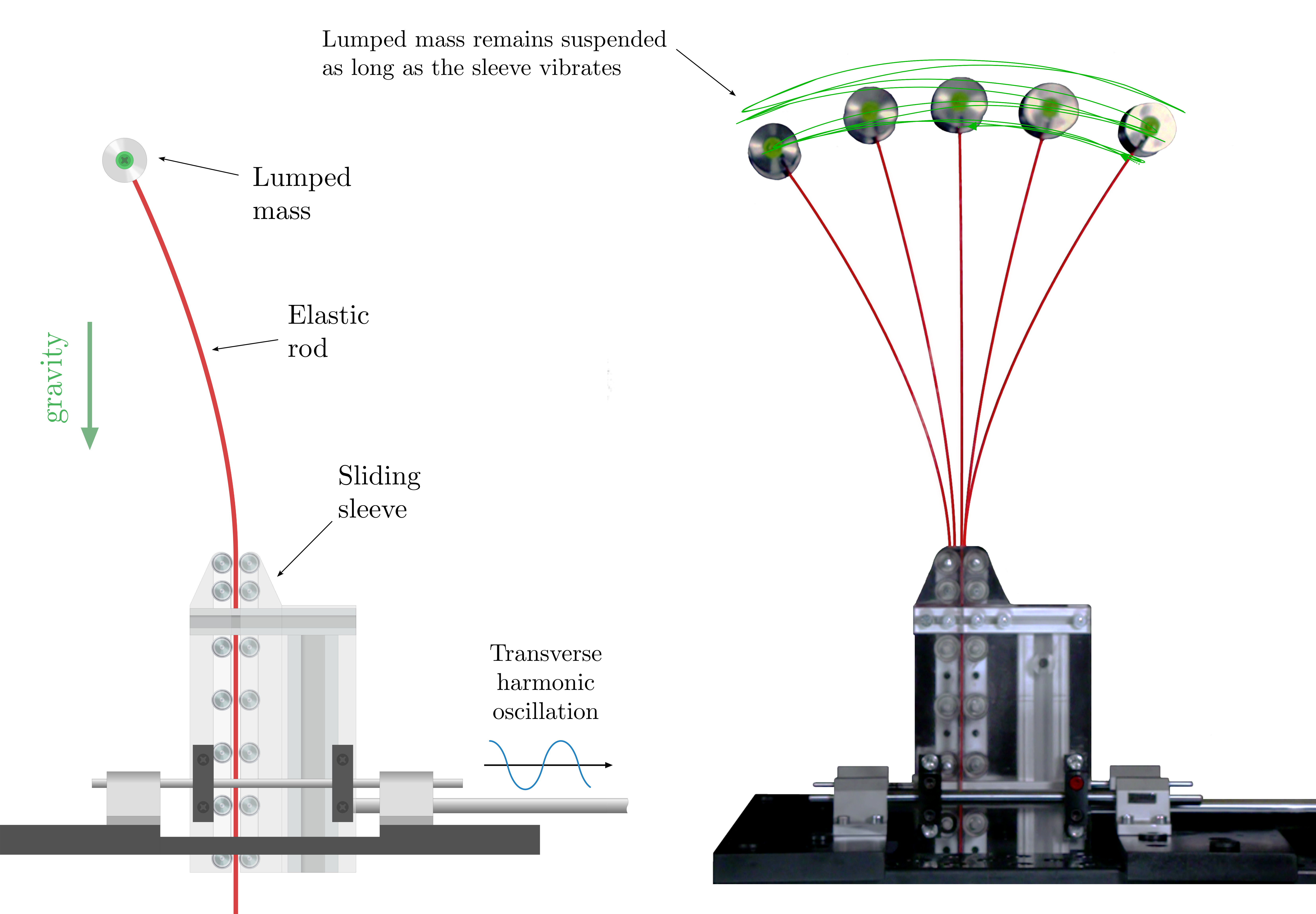}
        \caption{\footnotesize 
        (Left) Concept drawing of the experimental setup, where an elastic rod with a lumped mass attached at its top end is inserted inside a  sliding sleeve, with direction parallel to the gravity. While the mass would simply fall down  when the constraint does not move, the mass may oscillate suspended against gravity        
        when the constraint is  sinusoidally vibrating along  the horizontal direction.  (Right) 
        The suspended motion is shown through a  
        superimposed photo sequence and the tracked trajectory of the mass centre (green line). 
		}
        \label{sequenza}
    \end{center}
\end{figure}

The Kapitza inverted pendulum is a famous example of a 
parametric oscillator showing that  an unstable equilibrium configuration for a rigid and movable 
structure can be dynamically stabilized  through a controlled  time-harmonic vibration, acting at the basis of the pendulum and aligned parallel to the gravity \cite{kapit}. 
The aim of the present article is to disclose how  a similar, but more complex, phenomenon can be displayed by a variable length elastic structure through a controlled  time-harmonic vibration orthogonal to the gravity direction.
In particular, 
an {\it elastic flexible  rod} is arranged 
in a sort of inverse pendulum configuration (Fig. \ref{sequenza}, left), where the lower part is constrained by a (frictionless) sliding sleeve and the other end has a lumped mass  attached. In the presence of a gravitational field and in the absence of any disturbance, including any motion of the constraint, 
the  free fall motion would occur for the mass through the injection into the sliding sleeve
of the rod, remaining  undeformed. Therefore, contrary to the Kapitza pendulum, {\it the straight upward configuration 
never represents an equilibrium configuration}. 
Despite this difference, it is shown that a transverse time-harmonic oscillation of the sliding sleeve may generate a stable periodic or quasi-periodic motion around a finite value of  the external length of the rod. In practice, this is displayed as a   \lq  mass swinging suspended in the air', as experimentally demonstrated through the photo sequence  reported in Fig. \ref{sequenza} (right). Noteworthy, the imposed constraint vibration is {\it transverse} to the gravity direction, marking another difference with respect to the Kapitza pendulum,  where the vibration  is instead parallel.

The sustained motion  shown in Fig. \ref{sequenza} (right), which is theoretically, numerically, and experimentally demonstrated in the following sections, 
is made possible by the elastic flexibility of the rod, so that it would be impossible to be observed  in a rigid system such as the Kapitza pendulum. More specifically, when constraining a flexurally deformed rod, the sliding sleeve  generates a configurational force
contrasting gravity and proportional to the square of the bending moment at the sliding sleeve entrance. 
This configurational force was  
recently 
shown to act on 
elastic rods with variable length subject to bending 
\cite{bigoni2015eshelby, hanna2018partial, o2017modeling, o2015some, wang2022eshelbian}, torsion 
\cite{bigoni2014torsional}, and analyzed 
for buckling
\cite{liakou2018application, liakou2018constrained}. 
As a result, the configurational force represents the structural 
counterpart of 
the driving force  on defects, as conceptualized in 
Eshelbian mechanics
\cite{eshelby1999energy}; see also a recently provided mechanical interpretation
\cite{ballarini2016newtonian}.

The dynamics of the considered structure is characterized by two stable attractors   inherent to the presence of the sliding sleeve constraint and corresponding to  two possible final stages for the system, respectively associated with  a vanishing and an infinite value of the external length of the rod \cite{ARMANINI201982}.  
Indeed, the rod slips inside the sleeve when the configurational force is too low, while it is ejected  when  the force is too high.
In addition to these two inherent limit attractors, 
it is found that a third one may arise 
when a harmonic transverse oscillation of  the sleeve is imposed. 
This third attractor is associated with a finite value for the external length of the rod, so that the mass does not simply fall down, rather, it  remains suspended around a certain height. The height value depends on (i.)  the amplitude and frequency of the imposed 
harmonic motion, (ii.)  the bending stiffness of the rod, and 
(iii.)  the (lumped) mass value. Roughly speaking, the height value is approximately given by  the  external length of the clamped rod which would be at resonance under the same frequency of the harmonic motion
imposed to the sliding sleeve.  

Finally, the oscillating sliding sleeve system is shown to display {\it self-tuning properties}, consisting in the spontaneous adjustment of the  external length when  the amplitude and frequency oscillation parameters are \lq relatively slowly varying' as long as the third attractor is stable.

Attraction to a suspended non-equilibrium configuration has never been so far observed and is proven in this article through asymptotic solutions, numerical simulations, and the design, realization, and validation (at the {\it Instabilities Lab} of the University of Trento) of a new experimental setup permitting control and measure of the nonlinear dynamics of an elastic rod, subject to  large rotations and, simultaneously, to configurational changes of very large amounts. 

The presented  results introduce a new paradigm  in nonlinear structural dynamics and may find applications for  positioning tasks in  soft robotics \cite{alfalahi2020concentric,renda2021sliding,sipos2020longest}, as well as for wave mitigation devices \cite{lee2018mass,liu2015broadband,matlack2016composite,wang2012mechanically}
and environmental energy harvesters \cite{gibus2022high,ma2020acoustic,yu2020piezoelectric},  where self-tuning property can be exploited to improve  efficiency and to extend the frequency range of application.

The article is organized as follows: after the experimental evidence  
presented in Section \ref{sua-evidenza}, a nonlinear elastic model of the system is developed in Section \ref{modello}, together with an asymptotic solution. Numerical results from the integration of the nonlinear dynamics 
are provided in Section \ref{periodicaz}, showing the transition from  periodic to quasi-periodic motion. Finally, the presented results are validated in Section \ref{esperimentaz} through a comparison between theoretical predictions and experimental measurements.

\section{Experimental evidence of  sustained motion and self-tuning of  the oscillating rod}\label{sua-evidenza}

An experimental setup 
has been designed, manufactured, and tested (at the \lq Instabilities Lab' of the University of Trento) with the purpose of 
observing the dynamics 
of a rod inserted into a sliding sleeve, as sketched in Fig. \ref{sequenza} (left). A detailed presentation of the experimental setup is deferred to Section \ref{esperimentaz}.

The sliding sleeve is forced to  sinusoidal transverseoscillations along the horizontal direction
through the applied displacement  $u_g(t)$, defined by
\beq\label{uground}
    u_g(t)=\bar u_g \cos\left(2\pi \bar f \,t\right),
\eeq
where $\bar u_g$ and $\bar f$ are respectively the amplitude and the frequency of the sliding sleeve oscillation, while $t$ is the physical time.

\begin{figure}[!h]
    \renewcommand{\figurename}{\footnotesize{Figure}}
    \begin{center}
       \includegraphics[width=160mm]{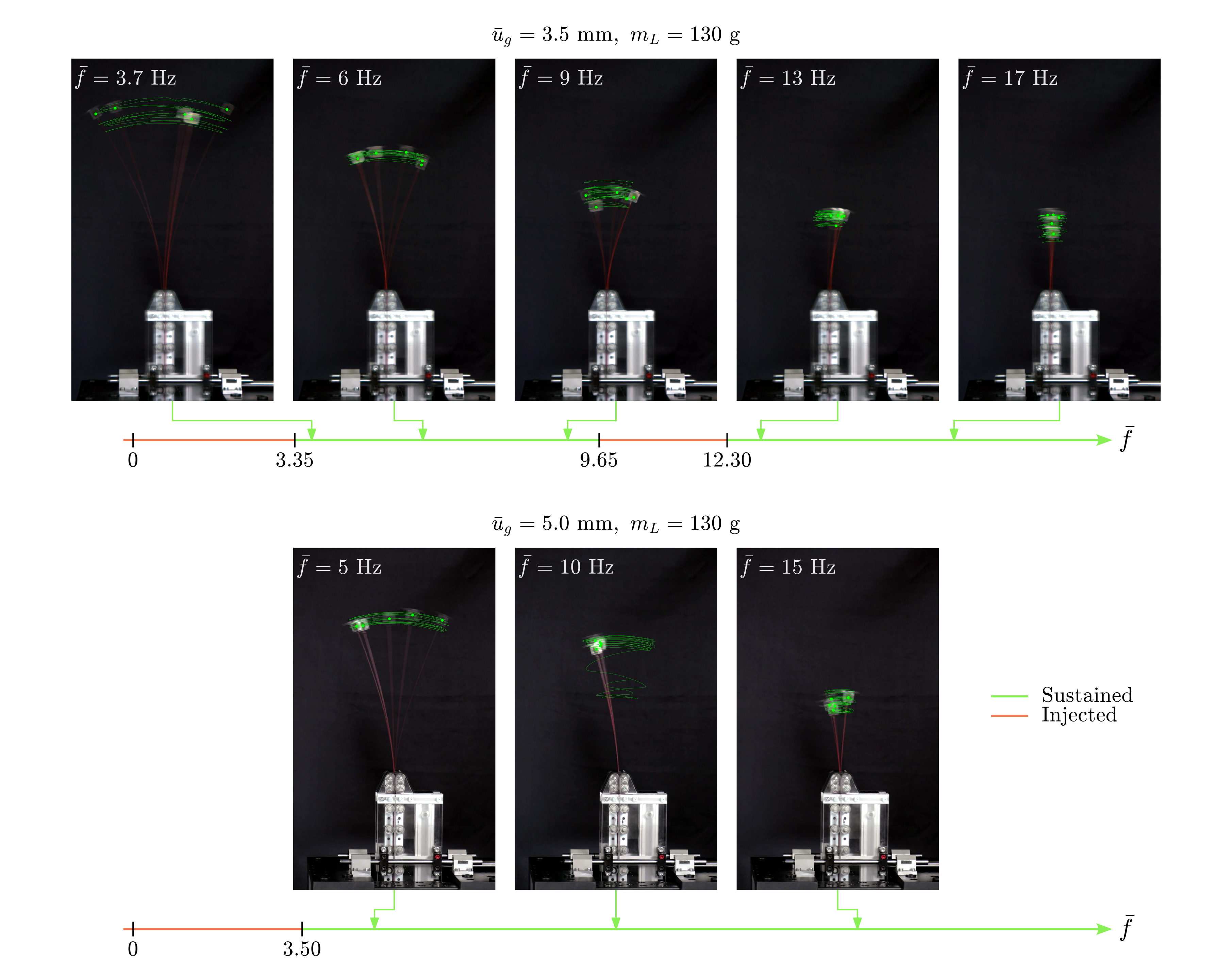}
        \caption{\footnotesize 
        A transverse vibration $u_g(t)=\bar u_g \cos\left(2\pi \bar f \,t\right)$ generates a suspended (against gravity) dynamic motion for a lumped mass $m_L=130$ g (tracked with green line) attached at the end of an elastic rod.
        Photos taken at different instants of time are superimposed for specific values of frequency $\bar f$, showing experimental results for two different oscillation amplitudes, $\bar u_g=3.5$ mm (top) and  $5$ mm (bottom). The green part of the $\bar f$ axis is associated with sustained oscillation, while the red one with final injection. 
       }
        \label{fig:sequenza_experiments}
    \end{center}
\end{figure}

\begin{figure}[!h]
    \renewcommand{\figurename}{\footnotesize{Figure}}
    \begin{center}
       \includegraphics[width=160mm]{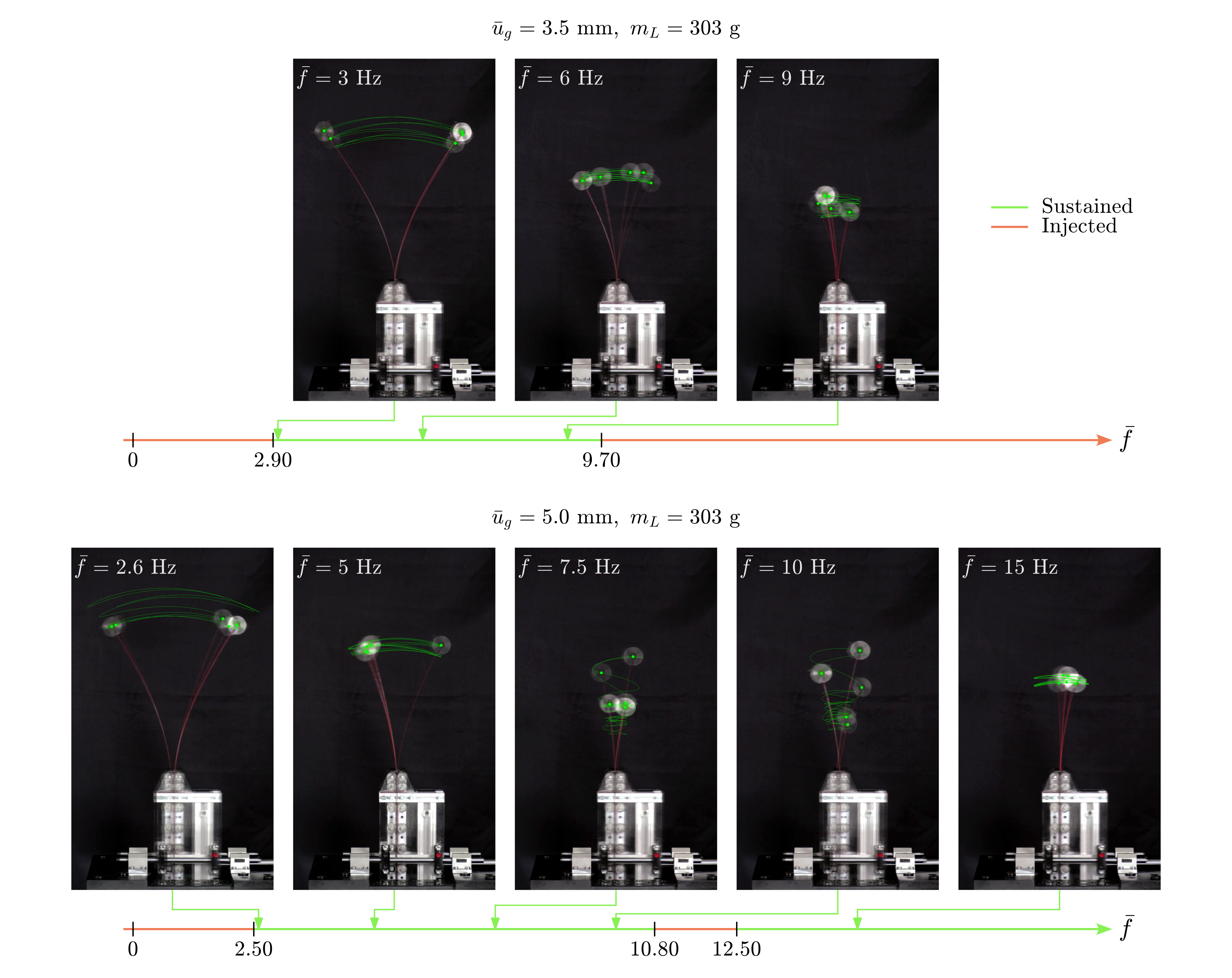}
        \caption{\footnotesize 
        As for Fig. \ref{fig:sequenza_experiments}, except for a lumped mass $m_L=303$ g. 
       }
        \label{fig:sequenza_experiments2}
    \end{center}
\end{figure}

When partially inserted inside of the vibrating sliding sleeve, the rod is intuitively expected to either fall down (at small oscillation amplitude or frequency) or to be ejected outside the sleeve (at high oscillation amplitude or frequency). These two behaviours do in fact  exist, but, interestingly, a third behaviour is found, in which the interplay of the configurational force generated at the sliding sleeve entrance with the gravity force creates a sustained motion, as shown in
Fig. \ref{sequenza} (right) and more accurately in Figs. \ref{fig:sequenza_experiments} and \ref{fig:sequenza_experiments2}, where a series of superimposed photos, with mass tracking reported with a green line, are shown at increasing frequency $\bar f$  for different lumped mass $m_L$ and amplitude $\bar u_g$ values. It can be appreciated that the sustained motion occurs for a height of the lumped mass varying with the frequency value, disclosing a self-tuning property of the system. Videos of the complete run of the experiments shown in  Figs. \ref{fig:sequenza_experiments} and \ref{fig:sequenza_experiments2} are available as Supplementary Material.

Since the beginning of the experimental campaign it has been observed that a sustained motion is not always displayed by the system, rather it can be achieved only within specific ranges of frequency and amplitude. With reference to the values of  the lumped mass $m_L=\{130,\,303\}$\,g, two kinds of experiments, both starting from a situation of sustained motion (held for 1 minute to overcome transient effects), have been performed:
\begin{enumerate}[(i)]
    \item \emph{amplitude controlled experiments} -  the frequency $\bar f$ is held  fixed at a specific constant value (ranging between 2.5 and 17\,Hz) while the amplitude $\bar u_{g}$ is decreased (with a step  of 0.1\,mm every 10 seconds) until injection, which terminates the experiment at a recorded final amplitude, denoted as \lq critical', $\bar u_{g,\,cr}(\bar f)$;
    \item \emph{frequency controlled experiments} -  the amplitude $\bar u_{g}$ is held fixed at a specific constant value ($\bar u_g=\{3.5,\,5\}$\,mm) while the frequency $\bar f$  is increased or decreased 
    (with a step of 0.05\,Hz every 15 seconds)
    until injection, which terminates the experiment at a recorded final frequency, denoted as \lq critical', $\bar f_{cr}(\bar u_g)$. 
\end{enumerate}

\begin{figure}[t]
    \centering
    \includegraphics[width=140mm]{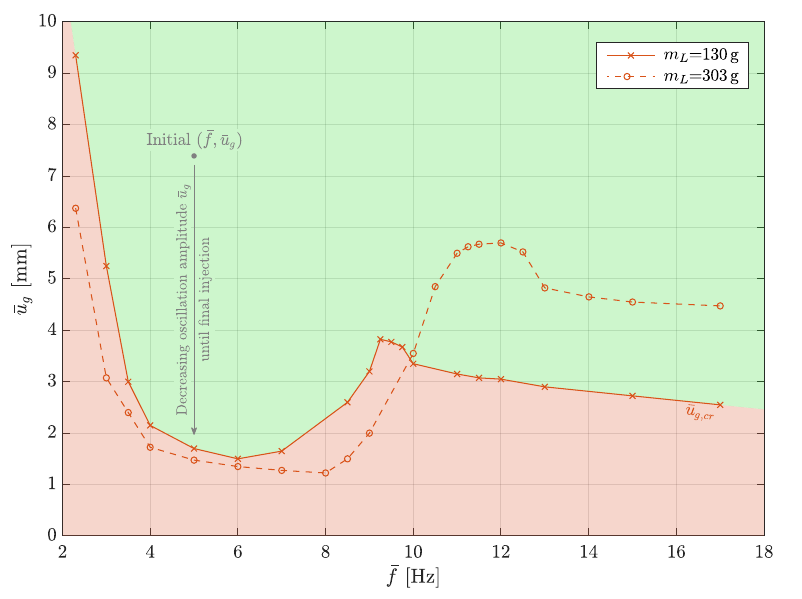}
    \caption{\footnotesize  Critical amplitude $\bar u_{g,\,cr}$ as a function of the frequency $\bar f$ (red curves). The critical curves are drawn through a linear interpolation of the critical pairs 
     measured through type (i) experiments for  $m_L$=130\,g (red crosses) and  for $m_L$=303\,g (red circles). 
     The critical curve defines the transition from a sustained motion (green region for  $m_L$=130\,g) to   the rod's final injection (red region for  $m_L$=130\,g).
    }
    \label{fig:crit_app}
\end{figure}

Experiments of the type (i) lead to the results
(in terms of 
critical amplitude $\bar u_{g,\,cr}$ 
versus frequency $\bar f$) reported in Fig. \ref{fig:crit_app}, 
where experimental measures for  the above-mentioned two different values of lumped mass correspond to different markers (crosses for $m_L=130$ g, and circles for $m_L=303$ g). The figure shows that the sustained motion is only possible for amplitude-frequency pairs belonging to the region above the critical curve (reported as green for $m_L=130$ g), while 
injection occurs  when the pair belongs 
to the region below the critical curve (reported red for  $m_L=130$ g).

At increasing frequency,  the results show an initial steep decrease of $\bar u_{g,\,cr}$, until a minimum is reached and a subsequent increase is visible, up to an isolated maximum, followed by a further mild decrease (within the measured frequency range).  
It follows that 
the sequence of tests of type (ii) experiments in order to detect the critical frequency $\bar f_{cr}$ may vary  from at least one to at least three, due to the presence of a non-unique $\bar f_{cr}$ for specific ranges of $\bar u_{g}$, as visible in the videos available as Supplementary Material. In this case, the critical values have to be detected  by  monotonically decreasing and  also monotonically increasing the frequency 
from different initial values. 
Results concerning the experiments of type (ii) are reported in Section \ref{esperimentaz}, where the developed theory is tested against the experiment to further validate the critical curves reported in Fig. \ref{fig:crit_app}.

\section{Mechanical modeling and average external length  }\label{modello}

\subsection{Nonlinear equations of motion}
\begin{figure}[!h]
    \renewcommand{\figurename}{\footnotesize{Figure}}
    \begin{center}
        \includegraphics[width=150mm]{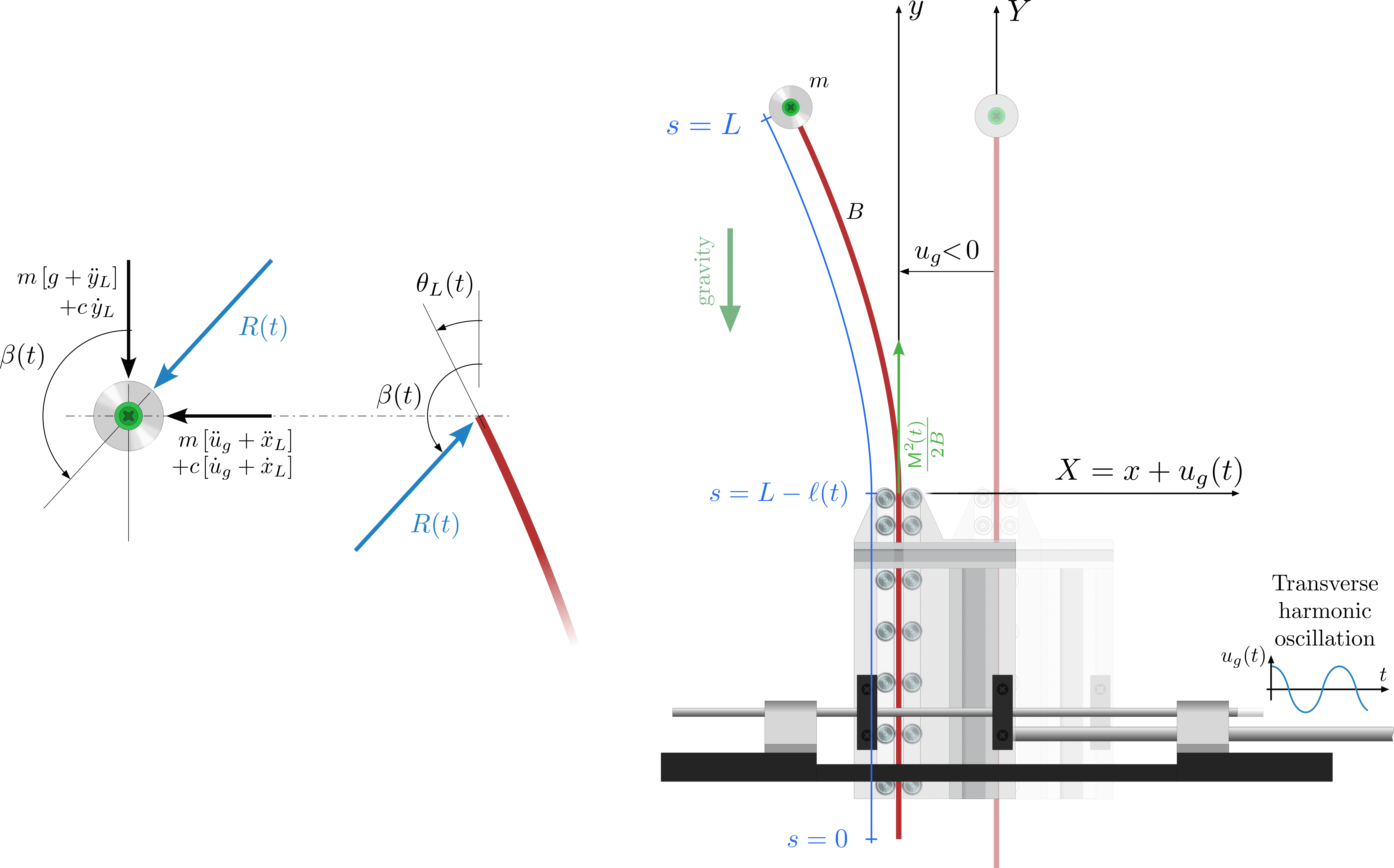}
        \caption{\footnotesize   An elastic rod, of flexural stiffness $B$ and total length $L$, is partially inserted into the sliding sleeve so that the time-varying external length is $\ell(t)<L$. A mass $m$ is attached to the upper end of the rod, while the sliding sleeve constraint horizontally oscillates in time through the displacement $u_g(t)$.
		}
        \label{sist}
    \end{center}
\end{figure}

The oscillating sliding-sleeve structure is modelled as an inextensible rod of length $L$, straight in its undeformed configuration,  with flexural stiffness $B$, and   partially inserted into a sliding sleeve, Fig. \ref{sist}. By introducing the curvilinear coordinate $s\in[0,L]$, corresponding to the arc-length of the rod's centre line, the portion of rod constrained by the sliding sleeve is defined by the set $s\in[0,L-\ell(t)]$, while the portion external to the sliding sleeve, which is the only part that can be bent, by $s\in[L-\ell(t),L]$, being $\ell(t)$ the  length of the rod external to the constraint. 
Note that the length $L$ of the rod is assumed to be finite, while the case of  infinite length would deserve a different treatment.
A mass $m$ is modelled as attached at the tip of the rod ($s=L$) and the sleeve is oriented in a vertical direction against gravity,  defined by the acceleration $g$. The attached mass value $m$ of the theoretical model is considered to be possibly different from the \lq real' lumped mass value $m_L$ in order to account for inertial effects of the rod (having a linear mass density $\gamma$), so that $m=m_L+\Delta m$, with  $\Delta m\geq0$. It is instrumental to consider two parallel Cartesian reference systems, the absolute system $X-Y$, with origin fixed at the mean position of the sliding sleeve entrance, and the relative system $x-y$, with origin attached at the moving sliding sleeve entrance and moving with it, which is forced to harmonically oscillate through the imposed displacement $u_g(t)$ (parallel to $X$), so that the absolute and the relative positions are related as
\begin{equation}
    X(s,t)=x(s,t)+u_g(t),
    \qquad
    Y(s,t)=y(s,t).
\end{equation}
Because of the rod inextensibility,  in addition to the configurational parameter $\ell(t)$, the deformed configuration is described through  the rotation field $\theta(s,t)$ measuring the counter-clockwise rotation of the rod's tangent with respect to the undeformed state. While the rotation is constrained to remain null within the portion of the rod inside the sliding sleeve, $\theta(s,t)=0$ for $s\in[0,L-\ell(t)]$, the rotation has to be evaluated outside the constraint. Neglecting the distributed inertia effects,  the governing equation for $\theta(s,t)$ is provided by the elastica  
\beq\label{geout}
\begin{array}{lll}
B\,\theta''(s,t)+N_x(t)\,\cos\theta(s,t)-N_y(t)\,\sin\theta(s,t)=0, \qquad
s\in[L-\ell(t),L],
\end{array}
\eeq
where the prime denotes differentiation with respect to $s$, $N_x(t)$ and $N_y(t)$ are the components along $x$ and $y$ axes of the internal action resultant $R(t)$, which is inclined at (the counter-clockwise) angle $\beta(t)$ with respect to the undeformed configuration,
\begin{equation}
    N_x(t)=R(t)\sin\beta(t),
    \qquad
    N_y(t)=-R(t)\cos\beta(t).
\end{equation} 
The differential equation (\ref{geout}) 
is complemented   by the boundary condition of null moment at the coordinate $s=L$ (neglecting  the rotational inertia of the  mass),
\beq
\theta'(L,t)=0,
\eeq
and the algebraic equation provided by the tangential force balance at the sliding sleeve entrance involving  the configurational force \cite{ARMANINI201982}
\beq\label{interfacial}
N_y(t) = -\frac{\mathsf{M}^2(t)}{2B},
\eeq
where $\mathsf{M}(t)$ is the bending moment at the sliding sleeve entrance, $\mathsf{M}(t)=B\,\theta'(L-\ell(t),t)$, which is given  by
\begin{equation}
\mathsf{M}(t)=N_x(t)\, y_L(t) - N_y(t)\, x_L(t),
\end{equation}
while $x_L(t)$  and $y_L(t)$ are the relative coordinates of the attached mass,
\begin{equation}
    x_L(t)=x(L,t),\qquad
    y_L(t)=y(L,t).
\end{equation}

It is highlighted that  Eq. \eqref{interfacial} is the approximated version  of the expression obtained in \cite{ARMANINI201982}, by neglecting  the possible contribution from the rotational inertia of the rod. The approximation 
presently introduced 
is motivated by the fact that the sliding velocity 
$\dot{\ell}(t)$
is much smaller than the longitudinal wave velocity in the rod. 

The internal force components can be obtained  through  D'Alembert's principle as 
\beq\label{boundcond}
\begin{aligned}
    N_x(t)=-m\left[\ddot x_L(t)+\ddot u_g(t)\right] - c(t) \left[\dot x_L(t) +\dot u_g(t)\right],
    \qquad
    &N_y(t)=-m\left[\ddot y_L(t)+g\right] - c(t) \,\dot y_L(t),
\end{aligned}
\eeq
where the overdot denotes differentiation with respect to the time $t$, and $c(t)$ is a linear time-variable viscosity coefficient defined as
\begin{equation}
    c(t) = 2\, \zeta\, \sqrt{\frac{3\, m\, B}{\ell^3(t)}},
\end{equation}
introduced to model the dissipation of the rod and the air drag effects outside the sliding sleeve through the non-dimensional parameter $\zeta\geq0$.

The spatial integration of the elastica (\ref{geout}) provides the  end's relative coordinates $x_L(t)$  and $y_L(t)$ and  the internal action components $N_x(t)$ and $N_y(t)$  as
\begin{equation}\label{elasticaintegration}
    \begin{array}{lll}
     & x_L(t) = \ell(t) \left\{ \mathcal{A}_1(t) \sin\beta(t) - \mathcal{A}_2(t)\cos\beta(t) \right\} ,\qquad
     & N_x(t) = \dfrac{B}{\ell^2(t)}\left[\mathcal{K}(k(t)) - \mathcal{K}(\sigma(t),k(t))\right]^2 \sin\beta(t), \\[2mm]
        & y_L(t) = \ell(t) \left\{ \mathcal{A}_1(t) \cos\beta(t) + \mathcal{A}_2(t)\sin\beta(t) \right\},\qquad
        & N_y(t) = -\dfrac{B}{\ell^2(t)}\left[\mathcal{K}(k(t)) - \mathcal{K}(\sigma(t),k(t))\right]^2 \cos\beta(t),
    \end{array}
\end{equation}
where $\mathcal{K}(k(t))$ and $\mathcal{K}(\sigma(t),k(t))$ are the complete and incomplete elliptic integrals of the first kind, respectively,
\begin{equation}
    \begin{array}{lll}
     & k(t) = \sin \left(\dfrac{\theta_L(t)-\beta(t)}{2}\right),
        & \mathcal{A}_1(t) = -1+\dfrac{2\left[\mathcal E(k(t)) - \mathcal E(\sigma(t),k(t))\right]}{\mathcal{K}(k(t)) - \mathcal{K}(\sigma(t),k(t))},\\[2mm]
        &\sigma(t) = -\mbox{arcsin} \left[ \dfrac{1}{k(t)}\, \sin\dfrac{\beta(t)}{2} \right],&\mathcal{A}_2(t) = -\dfrac{2\, k(t)\, \cos\sigma(t)}{\mathcal{K}(k(t)) - \mathcal{K}(\sigma(t),k(t))},
    \end{array}
\end{equation}
with $\theta_L(t)= \theta(L,t)$, while $\mathcal E(k)$ and $\mathcal E(\sigma,k)$ are the complete and incomplete elliptic integrals of the second kind.
Considering the results from the spatial integration (\ref{elasticaintegration}), the configurational force balance  (\ref{interfacial}) can be rewritten as
\begin{equation} \label{eq:oss_elastica2}
    \cos\beta(t) = 2 \left\{ \left[2 \left(\mathcal E(k(t)) - \mathcal E(\sigma(t),k(t))\right) - \left( \mathcal{K}(k(t)) - \mathcal{K}(\sigma(t),k(t)) \right)\right] \sin\beta(t)\,\cos\beta(t) + k(t)\,\cos\sigma(t) \cos \left(2\beta(t)\right)  \right\}^2 .
\end{equation}
The dynamics of the structural system is therefore governed by  a (DAE) differential-algebraic system of five equations composed by (i.) the two algebraic kinematic equations (\ref{elasticaintegration})$_1$ and (\ref{elasticaintegration})$_3$, (ii.) the configurational force balance equation (\ref{eq:oss_elastica2}), and (iii.) the two following  equations of motion 
\begin{equation} \label{eq:oss_elastica00}
\begin{cases}
    \begin{aligned}
        & \begin{multlined} m\, \left[\ddot u_g(t) + \ddot x_L(t)\right] + c(t)\, \left[\dot u_g(t) + \dot x_L(t)\right]- \frac{B}{\ell^2(t)}\left[\mathcal{K}(k(t)) - \mathcal{K}(\sigma(t),k(t))\right]^2\, \sin\beta(t) = 0, \end{multlined}\\
        & \begin{multlined} m \left[g + \ddot y_L(t)\right] + c(t)\, \dot y_L(t) -\frac{B}{\ell^2(t)}\left[\mathcal{K}(k(t)) - \mathcal{K}(\sigma(t),k(t))\right]^2\, \cos\beta(t) = 0
        .\end{multlined}
    \end{aligned}
\end{cases}
\end{equation}
The dynamic response can be solved for the five unknowns in time, 
$x_L(t)$, $y_L(t)$, $\ell(t)$, $\theta_L(t)$ and $\beta(t)$, for the respective given initial conditions, fixed  set of structural parameters $m$, $B$, $L$, $g$, $\zeta$, and prescribed sliding sleeve motion $u_g(t)$.

\subsection{Analytical periodic oscillation through asymptotic expansion}\label{asymptmain}

In order to obtain an analytical prediction for the dynamics of the structure subject to time-harmonic motion of the sliding sleeve (\ref{uground}), the differential system \eqref{eq:oss_elastica00}  governing  the relative coordinates $x_L(t)$ and $y_L(t)$ of the end of the rod is approximated through its asymptotic expansion for small end rotation $\theta_L(t)$  under the assumption of  null viscous dissipation  ($\zeta=0$) as
\begin{equation}\label{eom_inf_damp}
\begin{cases}
    \begin{aligned}
        &\begin{multlined}  m\, \ddot x_L(t) + \frac{3\, B}{\ell^3(t)}\, x_L(t) = m\, \omega^2\, \bar u_g\, \cos(\omega\, t), \end{multlined}\\[0.3em]
        &\begin{multlined}  m\, \left[g+\ddot y_L(t)\right] = \frac{\mathsf{M}^2(t)}{2 B},\end{multlined}
    \end{aligned}
    \end{cases}
\end{equation}
where $\omega=2\pi\bar f$ is the angular frequency.
Under small rotation $\theta_L(t)$  and by  assuming $\beta(t)\approx \pm\pi/2$ (corresponding to the condition $|N_y(t)|\ll|N_x(t)|$), the following second-order expansions hold
\begin{equation}
\mathsf{M}(t)=2\, \frac{B\, \theta_L(t)}{\ell(t)},\qquad
x_L(t) = \frac{2}{3}\,\ell(t)\,\theta_L(t), \qquad y_L(t) = \ell(t) \left[ 1 - \frac{4}{15} \,\theta_L^2(t) \right].
\end{equation}
Introducing a reference rod's length $\ell_m\in(0,L)$, defined as a mean value measured during a given interval of time (to be specified later), a  dimensionless time $\tau_m$ can be introduced as 
\begin{equation}
\tau_m = t \sqrt{\frac{g}{\ell_m}},
\end{equation}
together with other dimensionless quantities, namely, the dimensionless load $p_m$, the dimensionless angular frequency $\Omega_m$, the dimensionless oscillation amplitude $U_m$, the dimensionless external length $\lambda(\tau_m)$, and the attached mass dimensionless coordinates $\xi(\tau_m)$ and $\eta(\tau_m)$,
\begin{equation}
    \begin{aligned}\label{dimensionless}
        &
        p_{m} = \frac{m\, g\,\ell_m^{\,2}}{B}, ~~~
        \Omega_m = \omega\, \sqrt{\frac{\ell_m}{g}}, ~~~
        U_m = \frac{\bar u_g}{\ell_m},  ~~~      \lambda(\tau_m) = \frac{\ell(t)}{\ell_m}, ~~~
        \xi(\tau_m) = \frac{x_L(t)}{\ell_m}, ~~~ \eta(\tau_m) = \frac{y_L(t)}{\ell_m},
    \end{aligned}
\end{equation}
so that
the approximated equations of motion \eqref{eom_inf_damp}  can be rewritten in the following non-dimensional form 
\begin{equation} \label{eq:eqsmallnd_damp}
    \begin{cases}~
        \begin{aligned}
            & \begin{multlined} \accentset{\ast\ast}{\xi}(\tau_m) + \frac{3}{p_{m}\, \lambda^3(\tau_m)} \xi(\tau_m) = \Omega_m^2\, U_m \,\cos(\Omega_m \,\tau_m), \end{multlined}\\[0.3em]
            &\begin{multlined} 1+\accentset{\ast\ast}{\eta}(\tau_m)  - \frac{2\, \theta^2_L(\tau_m)}{p_{m}\, \lambda^2(\tau_m)}=0,\end{multlined}
        \end{aligned}
    \end{cases}
\end{equation}
where the superscript symbol $\ast$ stands for the derivative of the relevant quantity with respect to the dimensionless time $\tau_m$.
The end's rotation  $\theta_L(\tau_m)$ and   dimensionless external length $\lambda(\tau_m)$ are sought as the following periodic functions   in the dimensionless time $\tau_m$ 
\begin{equation}\label{period_sol_zeta}
    \theta_L(\tau_m) = \varepsilon_\theta \cos(\Omega_m\tau_m),\qquad \lambda(\tau_m) = 1-\varepsilon_\ell \cos(2\Omega_m\tau_m).
\end{equation}
where  $\varepsilon_\theta$ and $\varepsilon_\ell$   are the small dimensionless amplitudes of the rotation and of the external length oscillations.
By assuming the periodic response (\ref{period_sol_zeta}),
the dimensionless equations of motion \eqref{eq:eqsmallnd_damp} truncated at the smallest order reduce to
\begin{equation}\label{perteqns}
  \begin{cases}~
    \begin{aligned}
        & \begin{multlined}
2\left[6\, \varepsilon_\theta-p_m\,\Omega_m^2(3\, U_m+2\,\varepsilon_\theta)\right]\cos(\Omega_m\,\tau_m)
        - 3\, \varepsilon_\ell\left[2\, \varepsilon_\theta-p_m\,\Omega_m^2\,(3 \,U_m+8\,\varepsilon_\theta)\right]\cos (3\,\Omega_m\,\tau_m)\approx 0,\end{multlined}\\[0.3em]
        &\begin{multlined}
         15\left[p_m-\varepsilon_\theta^2+\dfrac{p_m\, \varepsilon_\ell^2\,(1-8\,\Omega_m^2)}{2}\right]
        -\left[15\, \varepsilon_\theta^2+30\, p_m\, \varepsilon_\ell-p_m\,\Omega_m^2\left(8\,\varepsilon_\theta^2+60\, \varepsilon_\ell\right)
        \right]\cos(2\,\Omega_m\,\tau_m) \approx 0. 
\end{multlined}
    \end{aligned}
    \end{cases}
\end{equation}

The annihilation of the  time-independent term in Eq. \eqref{perteqns}$_2$ leads to
\begin{equation}
p_m=\dfrac{\varepsilon_\theta^2}{1-4\,\varepsilon_\ell^2\,\Omega_m^2},\qquad
\mbox{or, equivalently,}\qquad
\varepsilon_\theta=\pm \sqrt{p_m}\ \sqrt{1-4\,\varepsilon_\ell^2\,\Omega_m^2},
\end{equation}
the latter showing that two twin solutions exist, one with the rotation $\theta_L(\tau_m)$ in phase ($\varepsilon_\theta>0$) and another in counter-phase ($\varepsilon_\theta<0$) with the imposed motion $u_g(t)$ of the sliding sleeve.

Assuming that $\varepsilon_\ell\,\Omega_m\ll1$, the last equation reduces to
\begin{equation}\label{pmepsilontheta}
p_m=\varepsilon_\theta^2, \qquad
\mbox{or, equivalently,}\qquad
 \varepsilon_\theta=\pm\sqrt{p_m},
\end{equation}
which in turn simplify Eqs. \eqref{perteqns} to
\begin{equation}\label{order2}
 \begin{cases}~
    \begin{aligned}
        & \begin{multlined}
    2\, \varepsilon_\theta\left[6-\varepsilon_\theta\,\Omega_m^2(3\, U_m+2\,\varepsilon_\theta)\right]\cos(\Omega_m\,\tau_m)
    - 3\, \varepsilon_\theta\,\varepsilon_\ell\left[2 -\varepsilon_\theta\,\Omega_m^2\,(3\, U_m+8\,\varepsilon_\theta)\right]\cos(3\,\Omega_m\,\tau_m)\approx 0,\end{multlined}\\[0.3em]
        &\begin{multlined}
     \varepsilon_\theta^2\,\left[15 +30\, \varepsilon_\ell-\Omega_m^2\left(8\,\varepsilon_\theta^2+60\, \varepsilon_\ell\right)
    \right]\cos(2\,\Omega_m\,\tau_m) \approx 0.
\end{multlined}
    \end{aligned}
    \end{cases}
\end{equation}

The expansion (\ref{order2}) shows that  equations of motion  are satisfied at the smallest order when
\begin{equation}
\begin{cases}
    \label{eq:omue}
    ~\begin{aligned}
    & 6-\varepsilon_\theta\,\Omega_m^2\left(3\, U_m+2\,\varepsilon_\theta\right)\approx 0,\\
    & 15 -\Omega_m^2\left(8\,\varepsilon_\theta^2+60\, \varepsilon_\ell\right)\approx 0,
    \end{aligned}
    \end{cases}
\end{equation}
whose  solution  can  be obtained under three specific regimes, differing in the   inequality ruling the orders of $\varepsilon_\ell$ and $\varepsilon_\theta$, namely:
$$ 
    \left|\varepsilon_\ell\right|\ll \left|\varepsilon_\theta\right|^2,
    \qquad \left|\varepsilon_\ell\right|\approx \left|\varepsilon_\theta\right|^2,\qquad\mbox{and}\qquad\left|\varepsilon_\ell\right|\gg \left|\varepsilon_\theta\right|^2.$$

Among the above three regimes, only the solution referred to the second one is presented below, because it was found to be the only one capable to predict  the stable response obtained from the direct time integration of the differential-algebraic system \eqref{elasticaintegration}$_1$, \eqref{elasticaintegration}$_3$, \eqref{eq:oss_elastica2}, and \eqref{eq:oss_elastica00} in the next Section. 
However, the solutions corresponding to the two remaining  regimes  are included for completeness in Appendix \ref{appendix_full_regimes}.

\begin{figure}[!h]
    \centering
    \includegraphics[width=\textwidth]{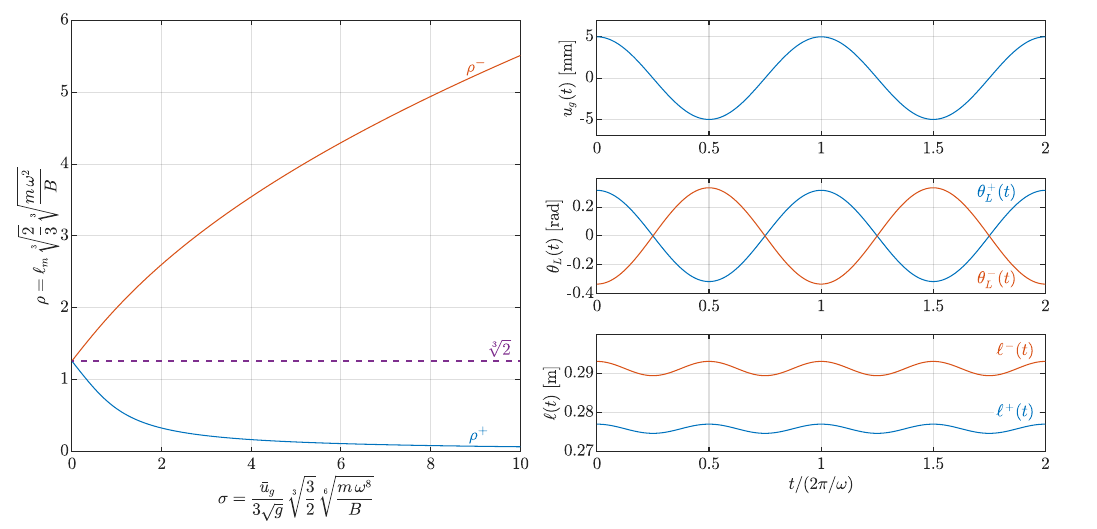}
    \caption{\footnotesize (Left part) Twin solutions $\rho^\pm$ versus parameter $\sigma$  (solid lines), Eqn. (\ref{ellm}). 
    In the limit of vanishing $\sigma$ , the two corresponding average lengths $\ell_m^\pm$ converge to the same value  $\ell_c=\sqrt[3]{3B/(m\omega^2)}$ (provided by  $\rho=\sqrt[3]{2}$, dashed line), defining the length of  the clamped rod at resonance with the constraint motion frequency. (Right part) Time-series of (top) the sliding sleeve harmonic motion $u_g(t)=\bar u_g\, \cos(\omega\, t)$ and the   twin periodic solutions in terms of (centre) the rotation $\theta_L^{\pm}(t)=\pm|\varepsilon_\theta|\cos(\omega t)$  and (bottom) the external length $\ell^{\pm}(t)=\ell_m^{\pm}[1-\varepsilon_\ell\cos(2\omega t)]$ for parameters $m=0.2$ kg, $B=1.5$ Nm$^2$, $\omega=10\pi$ rad/sec, and $\bar u_g=0.005$ m. The input motion $u_g(t)$ and the output motion $\theta(t)$ can be in phase or in counter-phase, each one corresponding to a different value of average external length $\ell_m$ but to the same dimensionless amplitude $\varepsilon_\ell$.}
    \label{fig:twin_sol}
\end{figure}

Under the assumption  $\left|\varepsilon_\ell\right|\approx \left|\varepsilon_\theta\right|^2$,  the system (\ref{eq:omue}) defines  the following asymptotic order for the dimensionless amplitude $U_m$ and angular frequency  $\Omega_m$
\begin{equation}\label{eq:UOm_eps_relation}
    U_m\approx \left|\varepsilon_\theta\right|,\qquad \Omega_m\approx \dfrac{1}{\left|\varepsilon_\theta\right|},
\end{equation}
implying the validity of the following solutions for $p_m\rightarrow 0$ and for finite values of $p_m \,\Omega_m^2$,
\begin{equation}\label{twin2}
    U_m = \frac{2 \left|3-p_m \,\Omega_m^2\right|}{3\, p_m\, \Omega_m ^2}\sqrt{p_m},\qquad
    \varepsilon_\ell=\frac{15-8\, p_m\,\Omega_m^2}{60\, p_m\, \Omega_m^2}\,p_m.
\end{equation}
A replacement in Eq. \eqref{twin2}$_1$ of the non-dimensional quantities $U_m$, $p_m$, and $\Omega_m$ with their dimensional counterparts (\ref{dimensionless})  
leads to the following twin equations
\begin{equation}\label{univcur_nodis}
    2\left(\frac{ m \,\omega ^2\,\ell_m^{\,3}}{B}-3\right) \pm 3\, \bar u_g\,\ell_m\,  \omega ^2 \,\sqrt{\frac{m }{g \, B}}=0, \qquad
    \mbox{for}\qquad \ell_m\neq\ell_{c}=\sqrt[3]{\frac{3B}{m\, \omega^2}},
\end{equation}
where $\ell_{c}$ represents the length of the clamped rod (with an attached  mass $m$ and bending stiffness $B$) at resonance with the sliding sleeve motion frequency $\omega$.
By setting two dimensionless positive parameters $\sigma$ and $\rho$, both unrestricted in their magnitude, as
\begin{equation}
\sigma = \frac{\bar u_g}{3\sqrt{g}}\,\sqrt[3]{\frac{3}{2}}\,\sqrt[6]{\frac{m \,\omega^8}{B}}= \frac{1}{3}\,\sqrt[3]{\frac{3}{2}}\,U_m\,\sqrt[6]{p_m}\,\sqrt[3]{\Omega_m^4}>0,\qquad
    \rho = \ell_m\,\sqrt[3]{\frac{2}{3}}\,\sqrt[3]{\frac{m\, \omega^2}{B}}=\sqrt[3]{\frac{2}{3}}\,
    \sqrt[3]{p_m\,\Omega_m^2}>0,
\end{equation}
 Eqs. \eqref{univcur_nodis} can be rewritten as the two following cubic dimensionless equations 
\begin{equation}\label{cubicone}
    \rho^3-2\pm 3\,\rho\,\sigma=0, \qquad\mbox{for}\,\,\rho\neq\sqrt[3]{2},
\end{equation}
each one admitting  a corresponding real and positive solution 
\begin{equation}\label{ellm}
    \rho^\pm(\sigma)=\dfrac{
    \sqrt[3]{
   \left(1+\sqrt{1\pm\sigma^3}\right)^2}
    \mp\sigma
    }
    {
    \sqrt[3]{
   1+\sqrt{1\pm\sigma^3}}
   },
\end{equation}
and reported as  functions of the dimensionless displacement amplitude $\sigma$ at the sliding sleeve entrance in  Fig. \ref{fig:twin_sol}\,(left). 
It is noted that in the limit case of vanishing values of $\sigma$ the two solutions $\rho^+$ and $\rho^-$ converge to the same value,
\begin{equation}
    \lim_{\sigma\rightarrow0}\rho^\pm(\sigma)=\sqrt[3]{2},
\end{equation}
and  therefore  both the  two average lengths $\ell_m^\pm$ tend to converge to  length  $\ell_{c}$ of the clamped system at resonance,
 \begin{equation}
\lim_{\frac{\bar u_g}{\sqrt{g}}\sqrt[6]{\frac{m \,\omega^8}{B}}\rightarrow0}\ell_m^\pm=\ell_{c}.
\end{equation}

The twin solutions $\theta_L^{\pm}(t)=\pm|\varepsilon_\theta|\cos(\omega \,t)$  and $\ell^{\,\pm}(t)=\ell_m^{\,\pm}[1-\varepsilon_\ell\cos(2\,\omega \,t)]$ are shown in Fig. \ref{fig:twin_sol} (right) for the parameters $m=0.2$ kg, $B=1.5$ Nm$^2$, $\omega=10\pi$ rad/sec, and $\bar u_g=0.005$ m.
Note that the rod's end rotation $\theta_L^+(t)$ is in phase with the sliding sleeve motion $u_g(t)$, while $\theta_L^-(t)$ is  in counter-phase, and that, although the twin solutions appear to be far from each other in Fig. \ref{fig:twin_sol} on the left, the difference in dimensional terms is small for the case under consideration ($\ell_m^-\approx 1.056 \ell_m^+$), as it can be appreciated from the graph of $\ell(t)$ (Fig. \ref{fig:twin_sol}, right bottom part).

It is interesting to note that the present asymptotic analysis shows that the periodic motion is a  solution  for the dynamic response  of  the structure in the limit case of  small rotations and high frequencies. Although based on these simplifying assumptions, it  provides a very good prediction for the average length $\ell_m$ of the part of the rod external to the sliding sleeve, as shown through the comparison with the numerical simulations in Sect. \ref{periodicaz} and with the experimental measurements in Sect. \ref{esperimentaz}. Moreover, the present solution reveals the self-tuning property of the system, disclosing how the average length $\ell_m$ of the periodic motion changes with varying  the oscillation amplitude $\bar u_g$ and angular frequency $\omega$.

\section{From periodic to quasi-periodic response}\label{periodicaz}

The asymptotic periodic solution presented in the previous section was derived with reference to the equations of motion approximated under the assumption of  small values of $\varepsilon_\theta$ and $\varepsilon_\ell$. As a result, the obtained analytical prediction  is relevant  when $|U_m|\sim 1/|\Omega_m| \to 0$, according to Eq. \eqref{eq:UOm_eps_relation}. Furthermore, the stability of such periodic response remains an open issue.
In order to provide a more complete view on the mechanical response and to assess the stability of the sustained motion, the nonlinear equations of motion \eqref{eq:oss_elastica00} 
are numerically integrated in time.

For the purposes of the numerical integration, the order of the differential equations \eqref{eq:oss_elastica00} is reduced by producing pairs of first-order differential equations. Then, the ordinary differential-algebraic equation system consisting of Eqs.\eqref{elasticaintegration}, \eqref{eq:oss_elastica2}, and the reduced-order version of Eqs. \eqref{eq:oss_elastica00} are discretized in time and the Crank-Nicolson method, adapted to ODEs, is used to proceed 
from one timestep to the next. The final algorithm was programmed using MATLAB (source code available as Supplementary Material). 
The obtained numerical predictions for the system response reveal that the mass  can display  three different types of motion at varying input parameters: (i.) periodic; (ii.) quasi-periodic; and (iii.) divergent, the latter leading either to a complete injection ($\ell(t)\rightarrow 0$) or ejection ($\ell(t)>L$) of the rod. The periodic and quasi-periodic vibrations represent a sustained motion around a finite value for the rod's external length, where the interplay between the gravitational  and configurational forces prevent the fall of the mass.

\begin{figure}[!h]
    \centering
    \includegraphics[width=190mm]{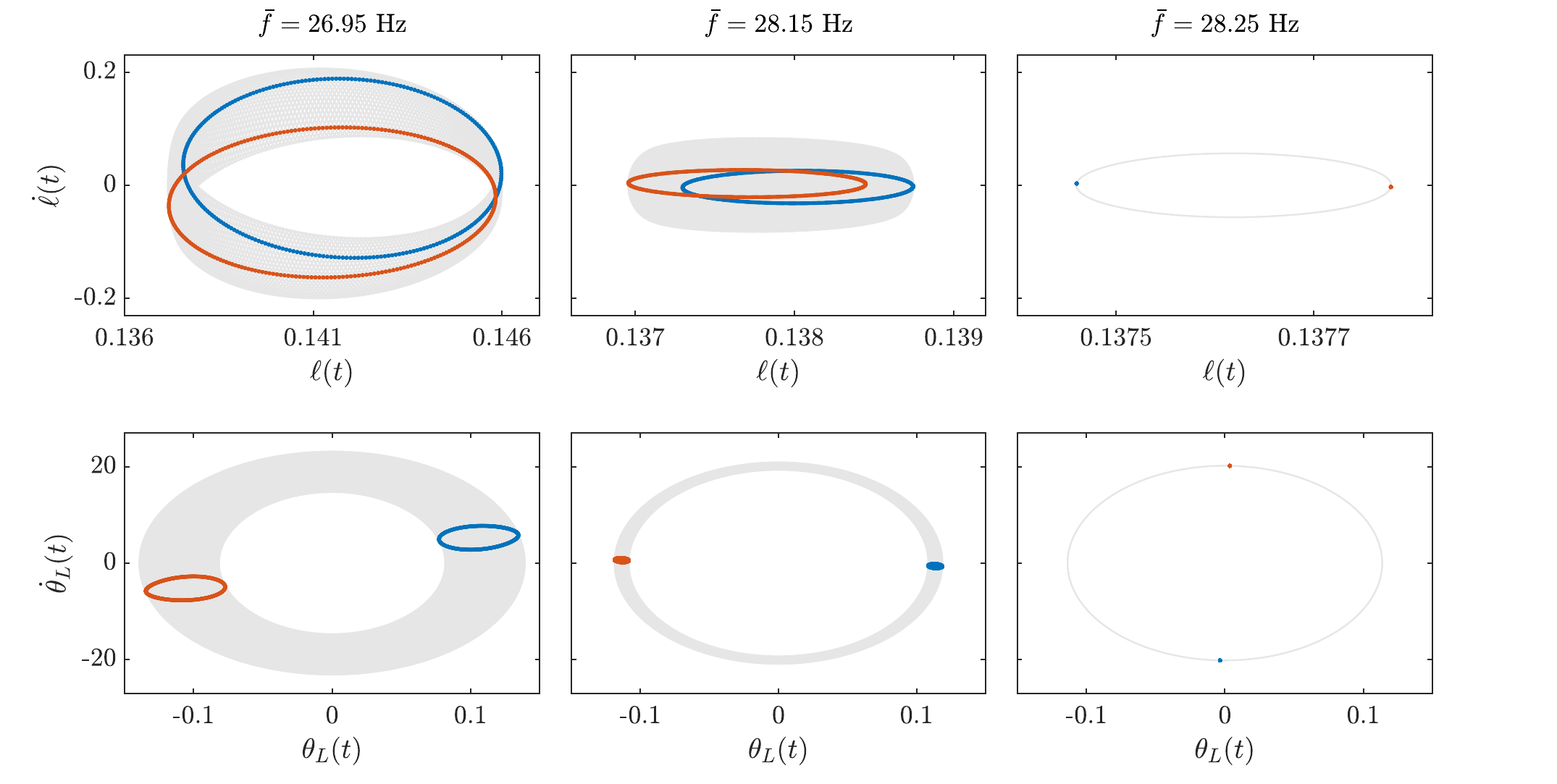}
    \caption{ \footnotesize Projections of the phase portraits and the Poincar\'e sections for the response of System $\mathsf{A}$, calculated by direct numerical integration of the equations of motion, for two quarter-period shifted phases of $\ell(t)$ (upper part) and two half-period shifted phases of $\theta(t)$ (lower part). Two qualitatively different  dynamic responses  are shown: (i.) quasi-periodic motion for excitation frequency $\bar f=26.95\mbox{Hz}$  (left) and  $\bar f=28.15\mbox{Hz}$ (centre) and (ii.) periodic motion for $\bar f=28.25 \mbox{Hz}$  (right), where the transition from the former to the latter occurs through a frequency decrease. The blue dots correspond to $t_1=k/\bar{f}$ and the red dots to $t_2=(k+1/2)/\bar f$ for the rotation $\theta_L(t)$ and to $t_2=(k+1/4)/\bar f$ for the external length $\ell(t)$. The grey regions denote the extent of the phase space covered in time by the system. 
    \label{fig:traj_high} }
\end{figure}

\begin{figure}[!h]
    \centering
    \includegraphics[width=190mm]{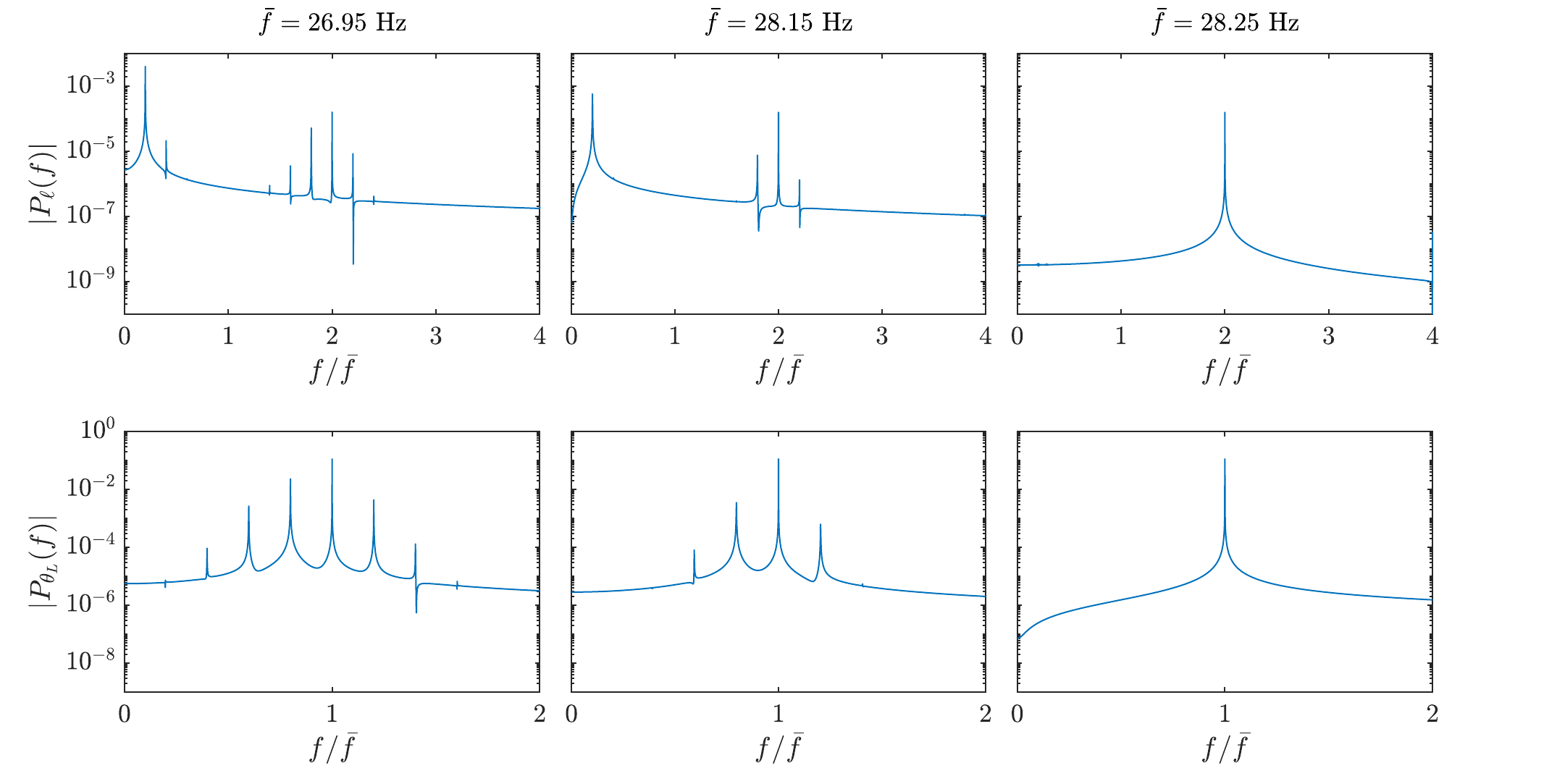}
    \caption{ \footnotesize As for Fig. \ref{fig:traj_high}, 
    except that 
    the spectrum of the response is analyzed, as calculated using the Fast Fourier Transform. 
    A single peak in the graphs occurring at rational values of $f/\bar{f}$ implies a periodic response, while peaks appearing for other values of $f/\bar{f}$ imply a quasi-periodic response.
    \label{fig:fft_high} }
\end{figure}

\begin{figure}[!h]
    \centering
    \includegraphics[width=\textwidth]{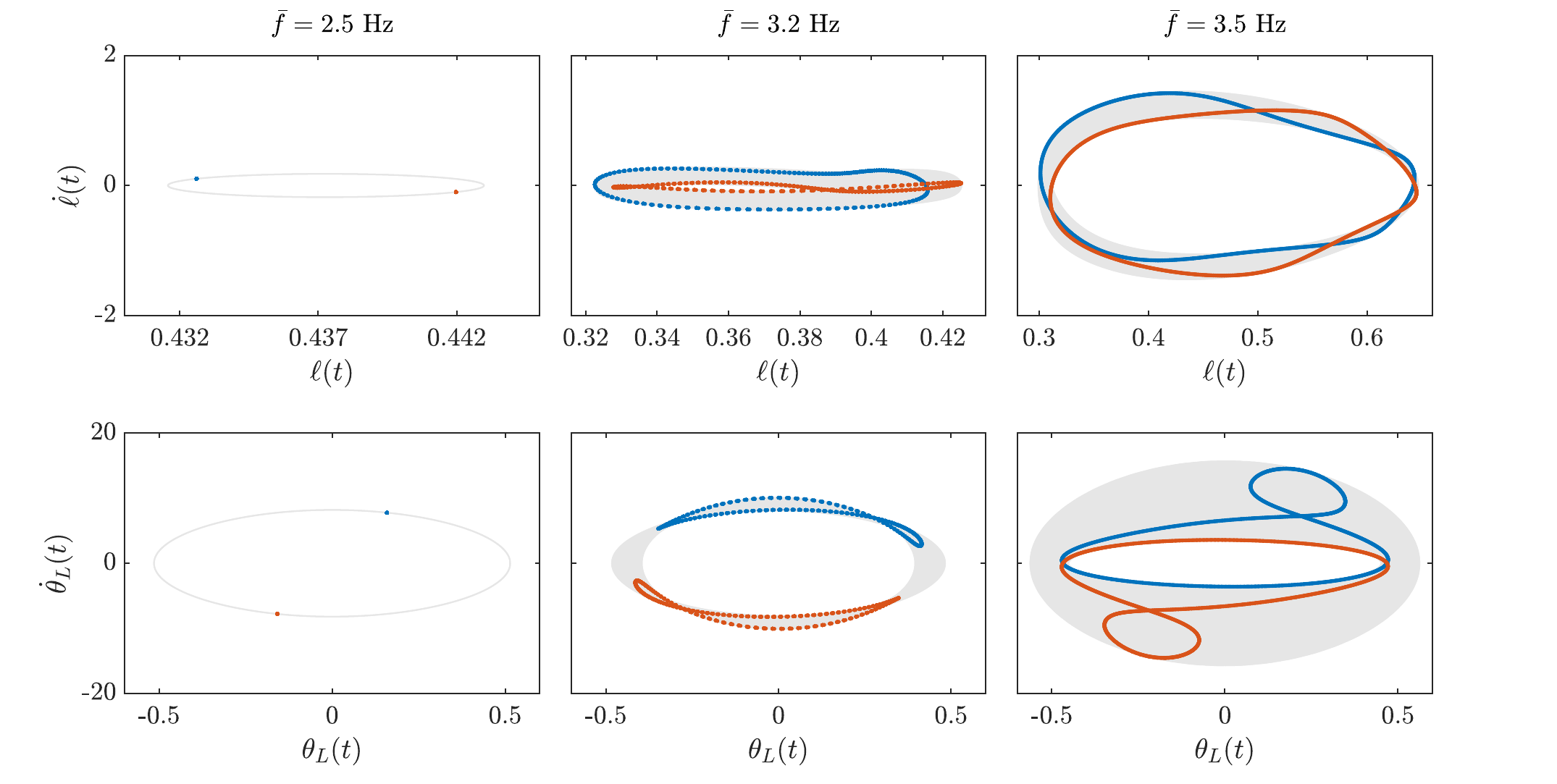}
    \caption{ \footnotesize 
    As for Fig. \ref{fig:traj_high}, 
    but for System $\mathsf{B}$ and showing that the transition from quasi-periodic ($\bar f=3.5 \mbox{Hz}$, right, and $\bar f=3.2\mbox{Hz}$, centre) to periodic ($\bar f=2.5\mbox{Hz}$, left) response can occur through a frequency decrease. \label{fig:oss_fft} }
\end{figure}

\begin{figure}[!h]
    \centering
    \includegraphics[width=\textwidth]{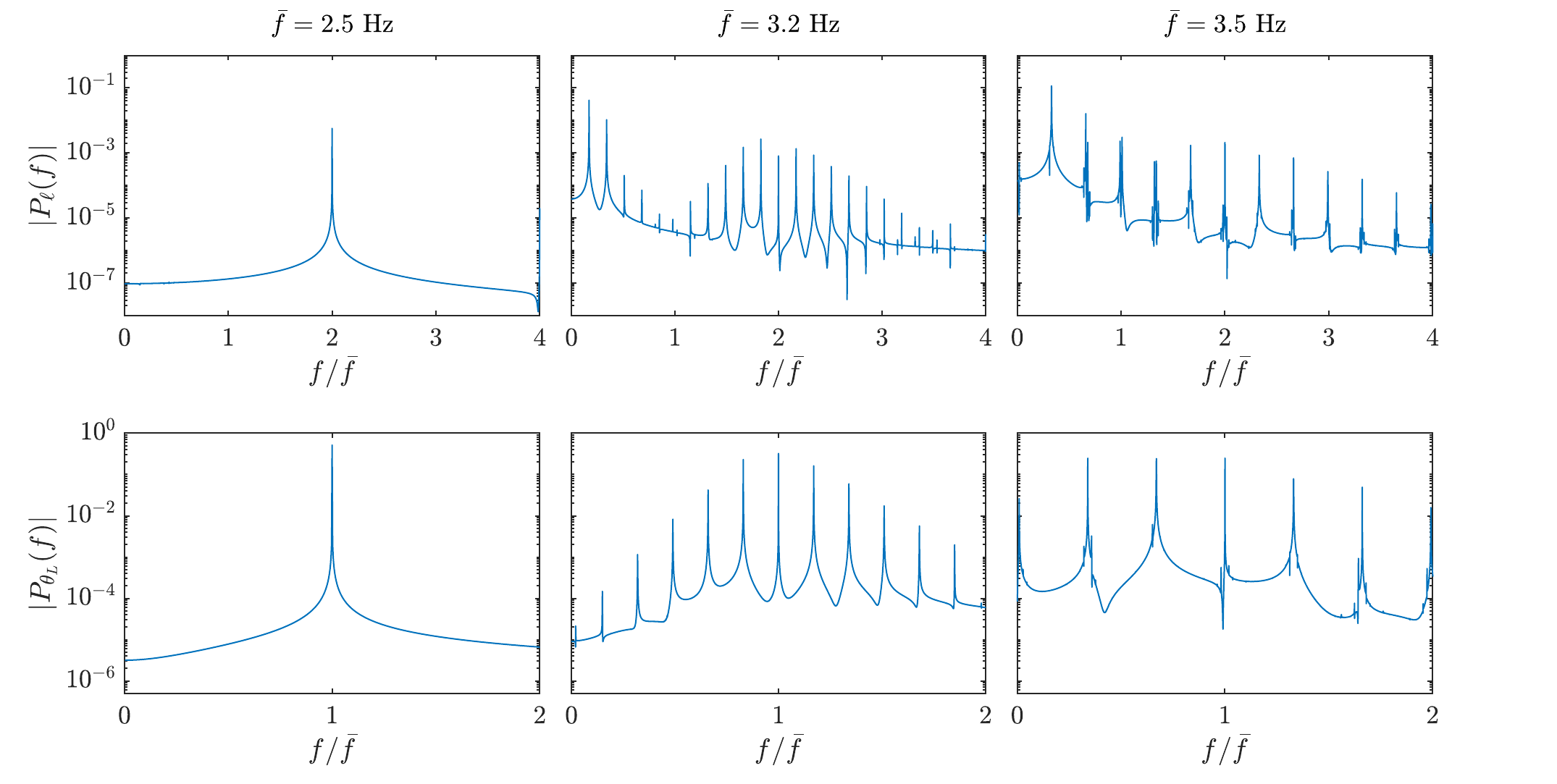}
    \caption{ \footnotesize As for Fig. \ref{fig:fft_high}, except that 
    the spectrum refers to System $\mathsf{B}$ as considered in Fig. \ref{fig:oss_fft}. \label{fig:oss_fft_theta} }
\end{figure}

By considering $B=1.4363\mbox{ Nm}$ and $ \bar u_g=5\mbox{ mm}$,  the two types of sustained motion are  shown in Figs. 
\ref{fig:traj_high} and \ref{fig:fft_high}  for System $\mathsf{A}$ (characterized by  $m=0.1\mbox{ g}$ and $\zeta=0.02$) 
and in Figs.  \ref{fig:oss_fft} and  \ref{fig:oss_fft_theta} for System $\mathsf{B}$ (characterized by $m=0.197\mbox{g}$ and $\zeta=0.005$, with the former value
 representative of the experimental setup corresponding to $m_L=130\mbox{ g}$, 
under the  
rod-lumped mass approximation, meaning that $m= m_L + \gamma L$). 
More specifically, the phase portraits of $\ell(t)$ and $\theta_L(t)$ are reported in Figs. \ref{fig:traj_high} and \ref{fig:oss_fft} together with their Poincar\'e sections, while the respective Fast Fourier Transforms $P_{\theta_L}(f)$ and $P_{\ell}(f)$  are reported in Figs. \ref{fig:fft_high} and  \ref{fig:oss_fft_theta}.
The results are reported when the transient effects due to the initial conditions on the motion are dissipated.

The Poincar\'e sections are shown for two shifted timings $t_1$ (blue dots) and $t_2$ (red dots), with $t_1=k/\bar f$ and $t_2=(k+1/2)/\bar f$ for the rotation $\theta_L(t)$ and with $t_1=k/\bar{f}$ and $t_2=(k+1/4)/\bar f$, for the external length $\ell(t)$, with $k\in\mathbb{N}$ defining the timings within the relevant interval. Moreover, the whole set of states occupied by the system is drawn as a grey region.
The Poincar\'e sections appear as a single point (and the whole set of states as an ellipse) when the response is periodic ($\bar f=28.25 \mbox{Hz}$ in Fig. \ref{fig:traj_high}, $\bar f=2.5\mbox{Hz}$ in Fig. \ref{fig:oss_fft}) or as a set of points resembling a closed loop  when the response is quasi-periodic ($\bar f=26.95\mbox{Hz}$  and  $\bar f=28.15\mbox{Hz}$ in Fig. \ref{fig:traj_high}, $\bar f=3.2\mbox{Hz}$ and $\bar f=3.5 \mbox{Hz}$ in Fig. \ref{fig:oss_fft}).
Therefore, it follows that the transition from periodic to quasi-periodic dynamics is shown in Figs. \ref{fig:traj_high} and \ref{fig:fft_high}   (Figs.  \ref{fig:oss_fft} and  \ref{fig:oss_fft_theta}) may occur at decreasing (or at increasing)   sliding sleeve oscillation frequency $\bar f$ for System $\mathsf{A}$ (System $\mathsf{B}$).

The change from periodic to  quasi-periodic can be also appreciated from the Fast Fourier Transforms  shown in Figs.  \ref{fig:fft_high}  and  \ref{fig:oss_fft_theta} through the appearance of  additional peaks in  correspondence to several frequencies different from $2 \bar f$ for  $P_{\ell}(f)$ and to $\bar f$ for $P_{\theta_L}(f)$. 

\begin{figure}[!h]
    \centering
    \includegraphics[width=\textwidth]{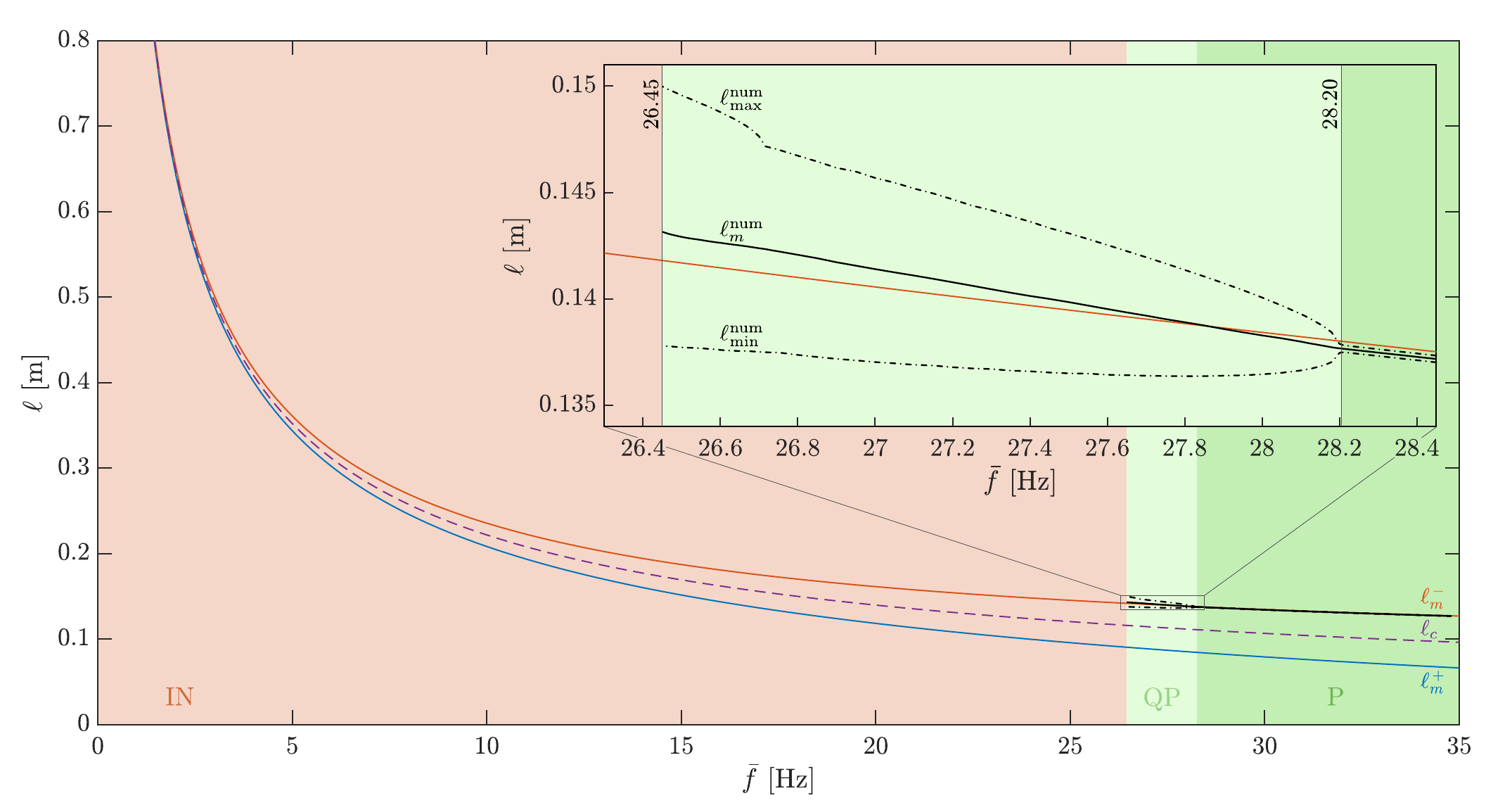}
    \caption{ \footnotesize Bifurcation diagram, showing the transition from periodic (green region, marked with \lq P') to quasi-periodic (light green region, marked with \lq QP') response of System $\mathsf{A}$, when the excitation frequency  is decreased to the value $\bar f\approx28.20$  Hz. When the frequency $\bar f$ is further reduced to reach  the  frequency $\bar f\approx 26.45$  Hz, the rod suffers injection (red region, marked with \lq IN'). The blue and red lines represent the two asymptotic evaluations for the average external lengths $\ell_m^{\,+}$ and $\ell_m^{\,-}$,  respectively, while the length $\ell_c$ of the clamped rod at resonance is reported as a dashed purple line.
The inset highlights the bifurcation and the transition points occurring at $\bar{f}=28.20$\,Hz and at $\bar{f}=26.45$\,Hz, respectively. 
\label{fig:oss_bifurcation} 
    }
\end{figure}

\begin{figure}[!h]
    \centering
    \includegraphics[width=\textwidth]{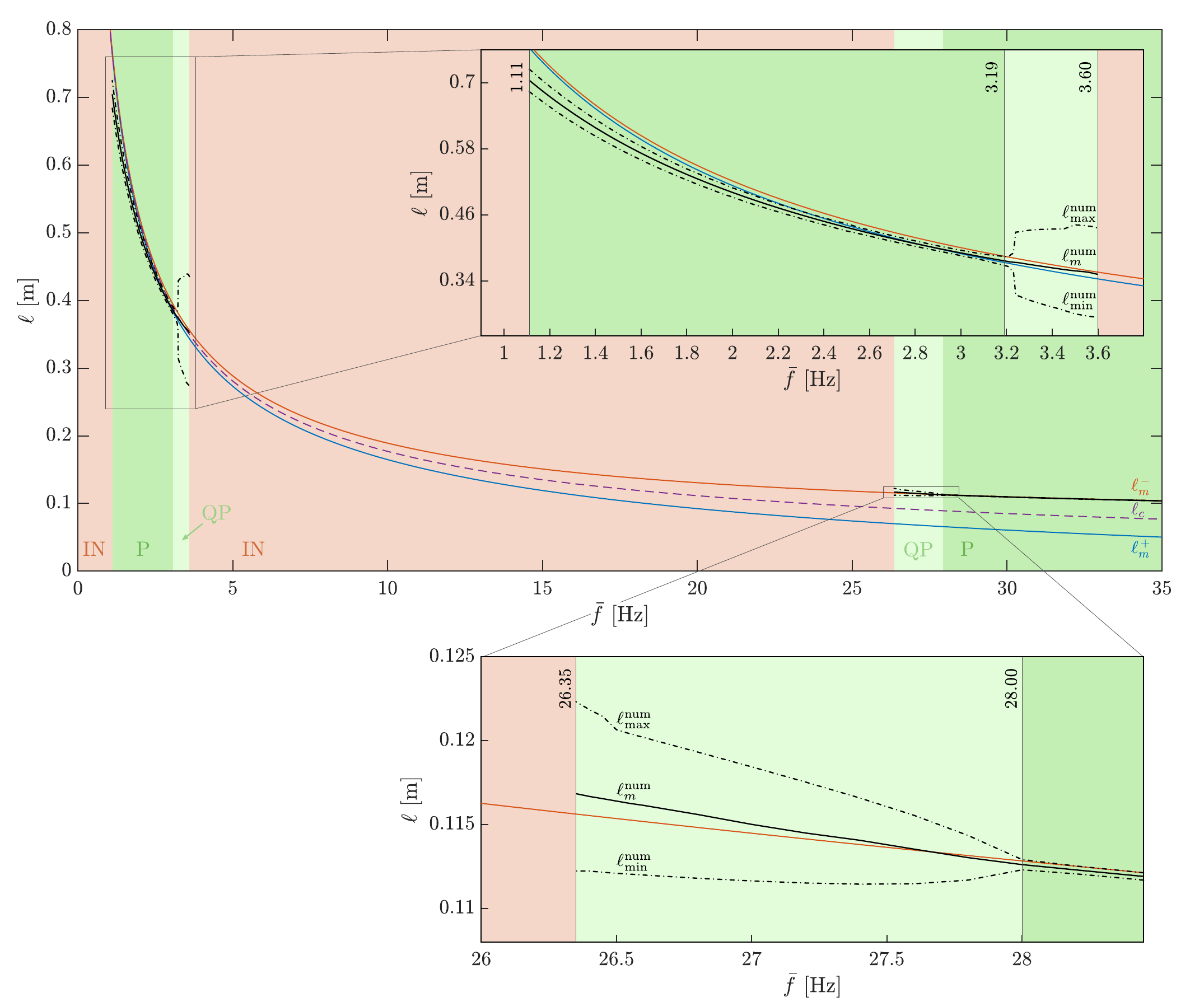}
    \caption{ \footnotesize As for Fig. \ref{fig:oss_bifurcation}, except that System $\mathsf{B}$ is analyzed. 
    Two ranges of frequencies are found (highlighted in the two insets) 
    within which the rod is subject to sustained motion. The loss of sustained motion can occur without displaying an intermediate quasi-periodic dynamics. \label{fig:oss_bifurcation2} }
\end{figure}

The conditions providing the transition between the different types of motion and when the sustained motion cannot occur are displayed over a wider set of frequencies through the diagrams reported in Figs. \ref{fig:oss_bifurcation} and \ref{fig:oss_bifurcation2} for Systems $\mathsf{A}$ and $\mathsf{B}$, respectively. In these diagrams, the numerical values of the average length $\ell_m^{\mbox{\scriptsize\hspace{.05em}num}}$ (solid black line)
and the corresponding  maximum $\ell_{\mbox{\scriptsize\hspace{.05em}max}}^{\mbox{\scriptsize\hspace{.05em}num}}$  and minimum $\ell_{\mbox{\scriptsize\hspace{.05em}min}}^{\mbox{\scriptsize\hspace{.05em}num}}$ length values  (dash-dotted black lines) are reported as functions of the frequency $\bar f$ and computed for each frequency value as
\begin{equation}
\ell_m^{\mbox{\scriptsize\hspace{.05em}num}}\left(\bar f\right)= \frac{\bar f}{k} \int_{t_0}^{t_0+k/\bar f} \ell(t)\,\mbox{d}t, 
\qquad
\ell_{\mbox{\scriptsize\hspace{.05em}max}}^{\mbox{\scriptsize\hspace{.05em}num}}\left(\bar f\right)= \max_{\substack{t\in [t_0, t_0+k/\bar f]}} \left\{ \ell(t) \right\}, 
\qquad
\ell_{\mbox{\scriptsize\hspace{.05em}min}}^{\mbox{\scriptsize\hspace{.05em}num}}\left(\bar f\right) = \min_{\substack{t\in [t_0, t_0+k/\bar f]}} \left\{ \ell(t) \right\},
\end{equation}
where $t_0$ is a time instant after the transient effects are dissipated and the integer $k$ (assumed equal to 60) is the number of oscillation periods used to determine $\ell_m^{\mbox{\scriptsize\hspace{.05em}num}}$, $\ell_{\mbox{\scriptsize\hspace{.05em}max}}^{\mbox{\scriptsize\hspace{.05em}num}}$, and $\ell_{\mbox{\scriptsize\hspace{.05em}min}}^{\mbox{\scriptsize\hspace{.05em}num}}$. 
The two average lengths $\ell_m^+$ (blue) and $\ell_m^-$ (red), analytically estimated through Eq. \eqref{ellm}, 
are also included in the diagrams, confirming how the asymptotic expression  $\ell_m^-$ provides an excellent estimation for the self-tuned external length $\ell$. 

Interestingly, the diagram shows that  the transition from periodicity (green region, marked with \lq P') to quasi-periodicity  (light green region, marked with \lq QP') is the result of a \emph{dynamic  bifurcation} of the system, where both $\ell_{\mbox{\scriptsize\hspace{.05em}min}}^{\mbox{\scriptsize\hspace{.05em}num}}$ and $\ell_{\mbox{\scriptsize\hspace{.05em}max}}^{\mbox{\scriptsize\hspace{.05em}num}}$ display a visible discontinuity in their derivative. Such dynamic bifurcation is followed by a monotonic increase in the differences  $\ell_{\mbox{\scriptsize\hspace{.05em}max}}^{\mbox{\scriptsize\hspace{.05em}num}}-\ell_m^{\mbox{\scriptsize\hspace{.05em}num}}$ and $\ell_m^{\mbox{\scriptsize\hspace{.05em}num}}-\ell_{\mbox{\scriptsize\hspace{.05em}min}}^{\mbox{\scriptsize\hspace{.05em}num}}$, eventually ending in the loss of quasi-periodicity due to the complete injection of the rod within the sliding sleeve (red region, marked with \lq IN'). Under this circumstance, due to the nonlinearities inherent to the increasing oscillation amplitude   of the rod's external length, the relation $ \left|\varepsilon_\ell\right|\approx \left|\varepsilon_\theta\right|^2$ does not hold anymore and the periodic solution \eqref{period_sol_zeta}, obtained through an asymptotic technique, becomes no longer representative of the structural response at varying frequency of the sliding sleeve. Moreover, it is also worth to mention that the loss of sustained motion may theoretically occur even without passing through a quasi-periodic response, as shown at low frequencies ($\bar f \approx$ 1 Hz) in Fig. \ref{fig:oss_bifurcation2}.
Overall, the diagrams show that the sustained motion occurs when  $\bar f> 26.45\mbox{ Hz}$ for System $\mathsf{A}$ and for  $1.11 \mbox{ Hz}<\bar f< 3.60 \mbox{ Hz}$ and $\bar f> 26.35\mbox{ Hz}$ for System $\mathsf{B}$. This  result 
confirms that, in agreement with the experimental observations in Sect. \ref{sua-evidenza} (Figs. \ref{fig:sequenza_experiments}--\ref{fig:crit_app}), the present structural model  does  display not only a (stable) self-tuned sustained motion for high frequencies but it  may also display it for an \lq island' of intermediate set of frequencies  depending on the elastic, inertial, and dissipation parameters.
Note that, while the present model based on linear damping never displayed  chaotic behaviour, the introduction (not pursued here)  of nonlinear dissipation, for instance friction, may deeply change the dynamics of the system, thus leading to chaos.

\section{Experimental setup and validation of the self-tuning external length value}\label{esperimentaz}

\begin{figure}[!h]
    \centering
    \includegraphics[width=140mm]{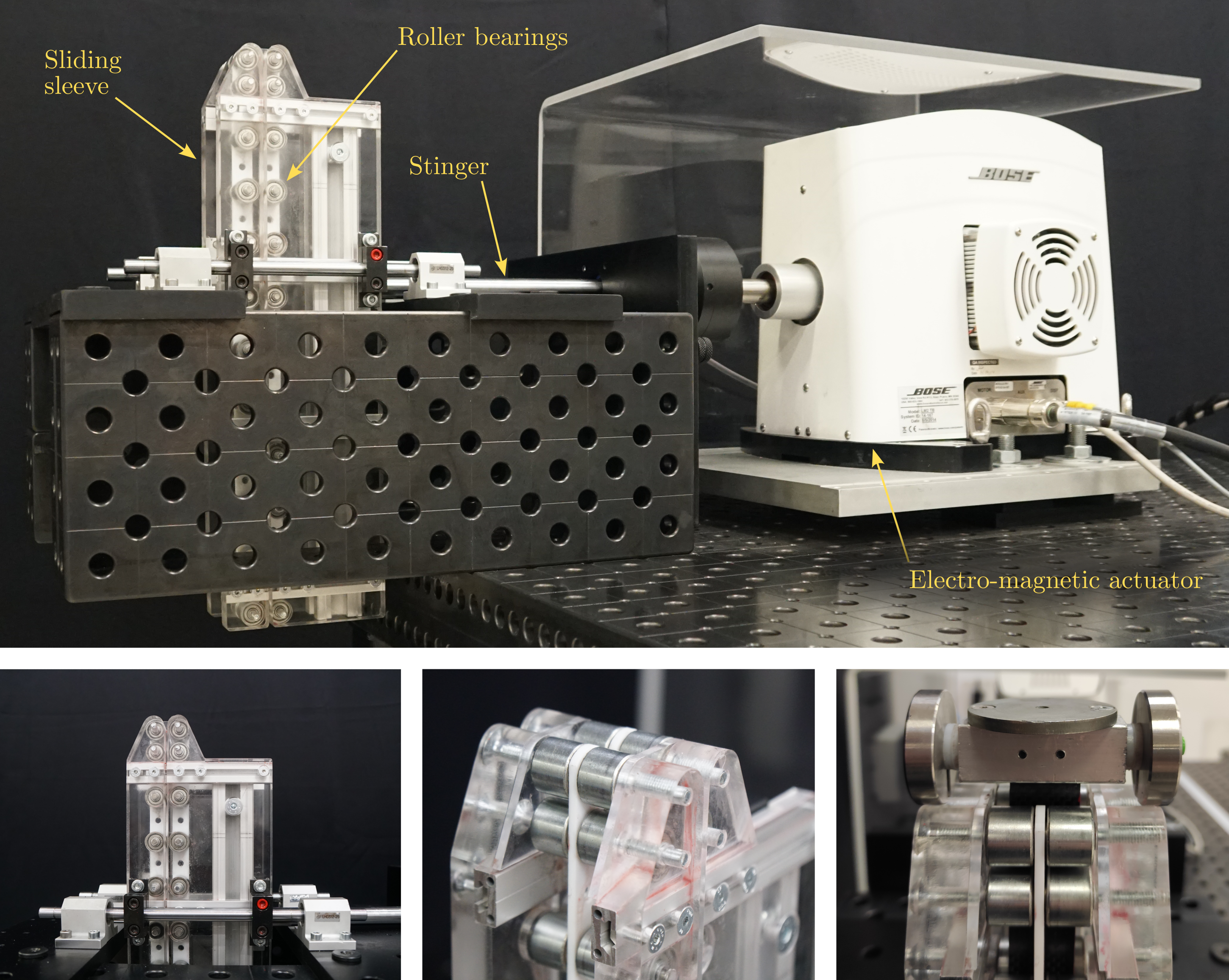}
    \caption{\footnotesize Experimental setup designed for the oscillating sliding-sleeve system (realized at the Instabilities Lab of the University of Trento). Upper part: general arrangement. Lower part: details of the sliding sleeve (left and centre) and of the lumped mass $m_L$ (right).} \label{fig:exp_setup}
\end{figure}

The 
experimental setup developed for testing the sliding rod is shown in Fig. \ref{fig:exp_setup} and was designed, manufactured and tested at the Instabilities Laboratory of the University of Trento.

The sliding sleeve is realized through two parallel arrays of rollers kept at a small fixed distance by means of two acrylic panels. More specifically, the distance between the roller arrays is set to maintain the constrained part of the rod straight, but still free to slide.
The sliding sleeve constraint device is connected through a stinger to an electromagnetic actuator (ElectroForce Linear motor 3300 Series II by Bose, frequency range 1-100 Hz, rated peak force SINE/RANDOM 3000N, max rated travel 25 mm) 
and screwed on a rail system that ensures motion along the horizontal direction only. Before each test, the sliding sleeve was oiled using Ballistol. 
 
 Carbon fiber rods are used in the experiments because of their modest material damping. The rods have a thin rectangular cross section and are realized by cutting a carbon-fiber sheet. The rod  properties are summarized in Table \ref{tab:rod}. 
 The lumped mass $m_L$ is made with a pair of interchangeable metallic disks  mounted at the rod's end.  Two values of the lumped mass $m_L$ were  selected, 130 and 303 g. The equipment was designed to allow the excitation frequency $\bar f$ and amplitude $\bar u_g$ to lie within the ranges $[0,\,20]\, \text{Hz}$ and $[0,\,10]\, \text{mm}$, respectively.
\begin{table}[!h]
    \caption{\label{tab:rod} Properties of the carbon-fiber rods used in the experiments.}
    \centering
    \begin{tabular}{cll}
        \toprule
        & Property & Value \\\midrule
           & Thickness & 2.04 mm \\
                                            & Width & 25.33 mm \\
                                            & Length $L$ & 800 mm \\
                                            & Bending stiffness $B$ & 1.4363 Nm$^2$ \\
                                            & Mass $(\gamma\,L)$ & 67 g \\
        \bottomrule
    \end{tabular}
\end{table}

The rod's dynamics was recorded during the experiments  with a Phantom High Speed V2640 camera and a Sony PXW-FS7 camera and the related videos were post-processed using a Matlab script in order to extract the time-histories of the relative coordinates, $x_L(t)$ and $y_L(t)$, of the mass and the external length of the rod $\ell(t)$.

\begin{figure}[!h]
    \centering
    \includegraphics[width=140mm]{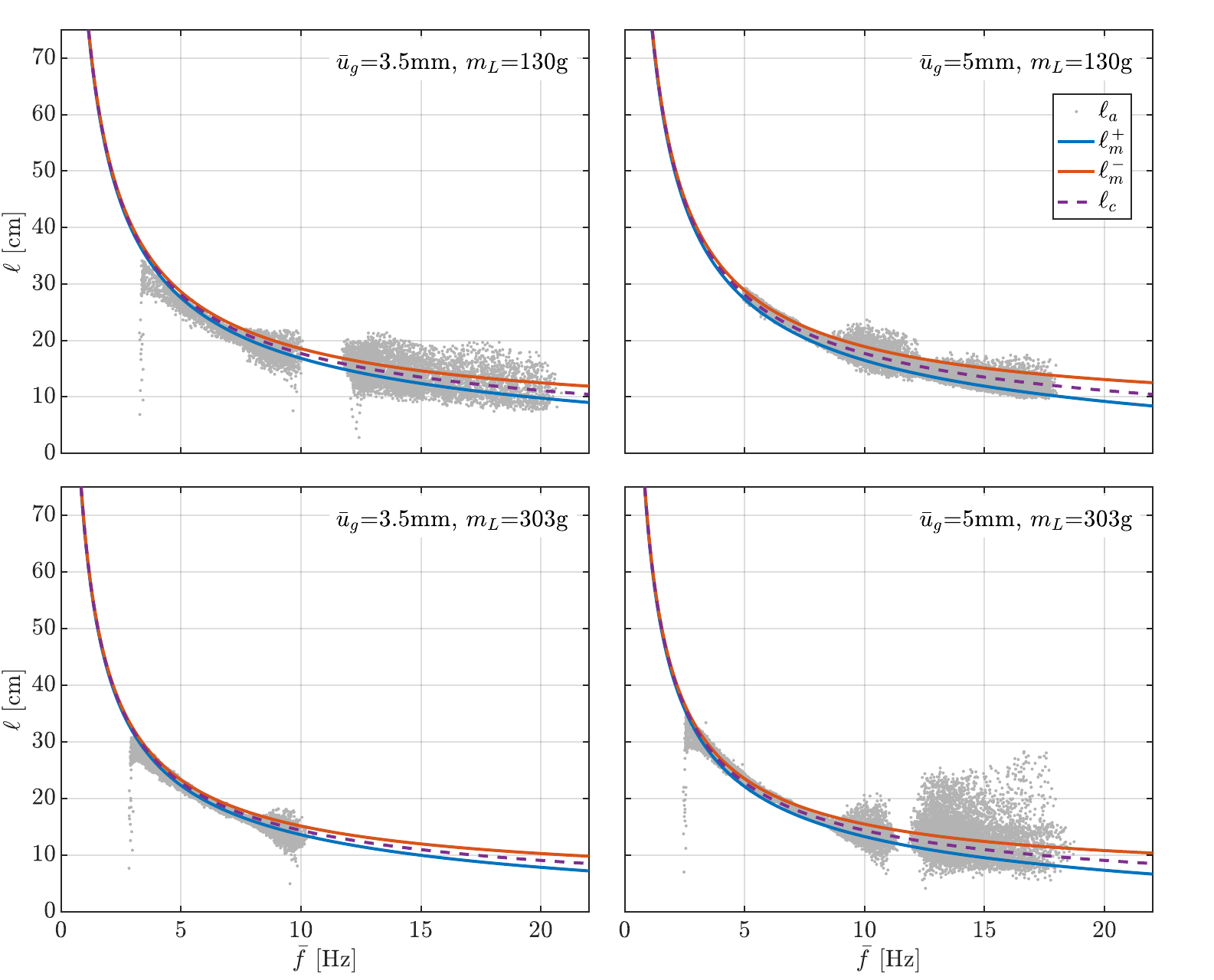}
    \caption{\footnotesize  External  length 
$\ell$ of the rod versus oscillation frequency $\bar f$
    for systems with lumped mass $m_L=\{130,303\}\,\text{g}$ and constraint amplitude $\bar  u_{g}=\{3.5,5\}\,\text{mm}$. The gray points correspond to the period-averaged external length $\ell_a^{(k)}$ evaluated from the  experimental measurements through Eq. (\ref{lak}) (data collection available as Supplementary Material). The blue and red solid lines respectively correspond to the asymptotic predictions $\ell_m^{\,+}$ and $\ell_m^{\,-}$,  Eq. \eqref{ellm}. while the dashed purple line represents the length $\ell_c$ of the clamped rod at resonance.  Absence of  stability for the sustained motion can be observed for an intermediate range of frequencies where a gap is observed between the grey dots (the absence of gray dots at higher frequencies is due to the lack of experimental data), resulting in a complete injection of the rod. When the sustained motion is stable, the system displays self-tuning property through its spontaneous adjustment in the value of the external length $\ell$ as a consequence of  a small variation in the oscillation frequency $\bar f$.}
    \label{fig:exp_len}
\end{figure}

Quantitative  experimental results are reported  in Fig. \ref{fig:exp_len} for every pair of values $\bar u_g$ and $m_L$ by performing the experiments of type (ii). In addition to the trajectories reported in Fig. \ref{fig:sequenza_experiments}, the experimental results are also shown in Fig. \ref{fig:exp_len} in terms of period-averaged external length $\ell_a^{(k)}$ at the $k$-th period of  the sliding sleeve oscillation, evaluated as
\begin{equation}\label{lak}
    \ell_a^{(k)}=\frac{1}{t_{k+1}-t_k}\int_{t_k}^{t_{k+1}} \ell (t) \mbox{d} t, \qquad k\in\mathbb{N},
\end{equation}
where $t_k$ ($t_{k+1}$) denotes the time instant at which the period begins (terminates). 

The experimentally evaluated values of $\ell_a^{(k)}$ appear as a cloud of gray dots (included as data collection and visible as a real-time collection in the videos  available as Supplementary Material), 
satisfactorily matching the asymptotically obtained average external lengths, $\ell_m^{\,+}$ and $\ell_m^{\,-}$,  evaluated (for the undissipated  system, $\zeta=0$) through Eq. \eqref{ellm} by considering  $m=m_L+\gamma\,L$.  
The average external lengths, $\ell_m^{\,+}$ and $\ell_m^{\,-}$, are obtained under the assumption $ \left|\varepsilon_\ell\right|\approx \left|\varepsilon_\theta\right|^2$ and are complemented by reporting the length of the resonant clamped-free rod $\ell_c$. It follows that a major result  of the experimental campaign is the validation of the relation between the self-tuning external rod's length $\ell$ and the prescribed frequency of the  sliding sleeve, $\bar f$, Fig. \ref{fig:exp_len}. 
In a rough sense, the external length of the rod $\ell$ is shown to self-adjust (in an average sense) approximately within the range $[\ell_m^{\,+},\ell_m^{\,-}]$, so that the rod mass system evolves to reach a configuration close to its resonant state and a sustained motion is realized. However, the sustained motion  can be terminated by the occurrence of an instability in which the rod trajectory degenerates  into a final injection. 

\section{Conclusions}

A special kind of elastic inverted pendulum, 
for which static equilibrium is impossible at small deflections, has been analyzed. 
 Interestingly, its inverted configuration can be maintained against gravity when a transverse  vibration is applied at the lower constraint. For specific ranges of system parameters the structure is shown (analytically, numerically, and experimentally) to self-tune its external length and  display a sustained motion, obtained through an interplay 
 between the downward (constant) gravity force and the upward (time-varying) configurational force, the latter strictly related to the elasticity of the structure. 

For the analyzed structure, the two following main features are found.
\begin{itemize}
    \item Existence of twin asymptotic periodic motions, around two different finite values of the average rod's length $\ell_m^{\,+}$ and $\ell_m^{\,-}$,  external to the sliding sleeve. These two values are shown to be close to $\ell_c$, corresponding to the length of a clamped rod at  resonance at  the input frequency.
Accurate numerical simulations,  accounting for a full nonlinear behaviour, have shown that only one ($\ell_m^{\,-}$) among the two periodic solutions may be stable.
    
    \item Stability of the sustained motion occurs only within a certain range of frequencies and amplitudes. Here, the sliding rod displays self-tuning properties by spontaneously adjusting its length as a result of the change in the oscillation amplitude and frequency.
\end{itemize}

The presented results may find application in compliant systems, 
including resonant metasources, where the self-tuning capability via   configurational constraints can be exploited for the development of wave mitigation devices and environmental energy harvesters based on flexible mechanisms. 
Moreover, the introduced mechanical concept can drive novel  control  applications where a longitudinal output can be tuned through a transverse input, with the purpose of chasing designed trajectories or to reach targeted positions.

\section*{Acknowledgements} 
PK  gratefully acknowledges the financial support from the European Union’s Horizon 2020 research and innovation programme under the Marie Sklodowska-Curie grant agreement ‘INSPIRE - Innovative ground interface concepts for structure protection’ PITN-GA-2019-813424-INSPIRE. DM, DB, and FDC  acknowledge financial
support from the European Research Council (ERC) under the European Union’s Horizon 2020 research and innovation programme
(Grant agreement No. ERC-ADG-2021-101052956-BEYOND).

\section*{References}
\printbibliography[heading=none]

\appendix

\section{Asymptotic solution under regimes for which $\left|\varepsilon_\ell\right|\not\approx \left|\varepsilon_\theta\right|^2$}\label{appendix_full_regimes}
The asymptotic solutions corresponding to the regimes $ \left|\varepsilon_\ell\right|\ll \left|\varepsilon_\theta\right|^2$ and $ \left|\varepsilon_\ell\right|\gg \left|\varepsilon_\theta\right|^2$ are reported.

\subsection{Regime  $ \left|\varepsilon_\ell\right|\gg \left|\varepsilon_\theta\right|^2$} 
In this regime the dimensionless amplitude $U_m$ and frequency $\Omega_m$ have the following asymptotic order
\begin{equation}
   \left|U_m\right|\gg \left|\varepsilon_\theta\right|,
   \qquad\left|\Omega_m\right|\approx \dfrac{1}{\sqrt{\left|\varepsilon_\ell\right|}},
\end{equation}
implying the following relations  at $p_m\rightarrow 0$ and for finite values of $\sqrt{p_m} \,\Omega_m^2$
\begin{equation}\label{Uandeps_1}
U_m =\pm\frac{2}{\sqrt{p_m}\Omega_m^2},
\qquad
\varepsilon_\ell=\frac{1}{4\sqrt{p_m} \Omega_m^2}\sqrt{p_m},
\end{equation}
the latter  showing that the solution defines $\varepsilon_\ell\approx \varepsilon_\theta$. Eq. \eqref{Uandeps_1} leads to the expressions for  the average length $\ell_m$ and its normalized amplitude $\varepsilon_\ell$ as
\begin{equation}
    \ell_m=\pm \dfrac{2}{\omega^2 \bar u_g}\sqrt{\dfrac{B}{m}},\qquad
    \varepsilon_\ell=\pm\dfrac{\bar u_g}{8}\sqrt{\dfrac{m g}{B}},
 \end{equation}
defining a gravity-insensitive average length $\ell_m$.

\subsection{Regime  $ \left|\varepsilon_\ell\right|\ll \left|\varepsilon_\theta\right|^2$} 
While in this regime the value $\varepsilon_\ell$ remains undefined, the dimensionless amplitude $U_m$ and frequency $\Omega_m$ have the following asymptotic order
\begin{equation}
    \left|U_m\right|\approx \left|\varepsilon_\theta\right|,\qquad \left|\Omega_m\right|\approx \dfrac{1}{\left|\varepsilon_\theta\right|},
\end{equation}
implying that the solutions valid at $p_m\rightarrow 0$ and for finite values of $p_m \,\Omega_m^2$ define
\begin{equation}\label{twin3}
    U_m = \pm\frac{2 \sqrt{p_m}}{5},\qquad
    \Omega_m=\frac{1}{2}\sqrt{\frac{15}{2 p_m}},
\end{equation}
which can be rewritten in terms of the average length $\ell_m$ and the oscillation amplitude $\bar u_g$ as
\begin{equation}
    \ell_m=\frac{1}{2}\sqrt[3]{\frac{15 B}{m \omega^2}},
    \qquad
   \bar u_g=\frac{1}{2}\sqrt[3]{\frac{9}{5\omega^4}}\sqrt[6]{\frac{B}{m}}\sqrt{g}.
\end{equation}

\end{document}